\catcode`\@=11

\newif\if@fewtab\@fewtabtrue

{\count255=\time\divide\count255 by 60
\xdef\hourmin{\number\count255}
\multiply\count255 by-60\advance\count255 by\time
\xdef\hourmin{\hourmin:\ifnum\count255<10 0\fi\the\count255}}
\def\ps@draft{\let\@mkboth\@gobbletwo
    \def\@oddhead{}
    \def\@oddfoot
       {\hbox to 7 cm{$\scriptstyle Draft\ version:\ \draftdate$
       \hfil}\hskip -7cm\hfil\rm\thepage \hfil}
    \def\@evenhead{}\let\@evenfoot\@oddfoot}

\def\ceqno{\global\@fewtabfalse
    \ifcase\@eqcnt \def\@tempa{& & &}\or \def\@tempa{& &}
      \or \def\@tempa{&}
      \or\def\@tempa{}\fi\@tempa
{\rm(\theequation)}}

\def\aeqno#1{\global\@fewtabfalse
    \ifcase\@eqcnt \def\@tempa{& & &}\or \def\@tempa{& &}
      \or \def\@tempa{&}
      \or\def\@tempa{}\fi\@tempa
{\rm(\theequation,#1)}}

\def\label#1{\ifnum\draftcontrol=1
 \global\def\draftnote{$\scriptstyle #1$}\fi
 \@bsphack\if@filesw {\let\thepage\relax
   \def\protect{\noexpand\noexpand\noexpand}%
\xdef\@gtempa{\write\@auxout{\string
      \newlabel{#1}{{\@currentlabel}{\thepage}}}}}\@gtempa
   \if@nobreak \ifvmode\nobreak\fi\fi\fi
  \@esphack}

\def\alabel#1#2{\label{#1}\global\@fewtabfalse
    \ifcase\@eqcnt \def\@tempa{& & &}\or \def\@tempa{& &}
      \or \def\@tempa{&}
      \or\def\@tempa{}\fi\@tempa
{\hbox to 3cm{\phantom{\rm(\theequation,#2)}
\draftnote \hfil}\hskip -3cm {\rm(\theequation,#2)}}}

\def\clabel#1{\label{#1}\global\@fewtabfalse
    \ifcase\@eqcnt \def\@tempa{& & &}\or \def\@tempa{& &}
      \or \def\@tempa{&}
      \or\def\@tempa{}\fi\@tempa
{\hbox to 3cm{\phantom{\rm(\theequation)}
\draftnote \hfil}\hskip -3cm{\rm(\theequation)}}}

\def\eqnarray{\def\draftnote{{}}\global\@fewtabtrue
\stepcounter{equation}\let\@currentlabel=\theequation
\global\@eqnswtrue
\global\@eqcnt\z@\tabskip\@centering\let\\=\@eqncr
$$\halign to \displaywidth\bgroup\@eqnsel\hskip\@centering\@eqcnt\z@
  $\displaystyle\tabskip\z@{##}$&\global\@eqcnt\@ne
  \hskip 1\arraycolsep \hfil${##}$\hfil
  &\global\@eqcnt\tw@ \hskip 1\arraycolsep
$\displaystyle\tabskip\z@{##}$
\hfil  \tabskip\@centering&\global\@eqcnt\thr@@\llap{##}\tabskip\z@
\cr}

\def\endeqnarray{\@@eqncr\egroup
      \global\advance\c@equation\m@ne$$\global\@ignoretrue}

\def\@eqnnum{\hbox to 3cm{\phantom{\rm(\theequation)} \draftnote
                         \hfil}\hskip -3cm {\rm(\theequation)}}

\def\@@eqncr{\let\@tempa\relax
    \ifcase\@eqcnt \def\@tempa{& & &}\or \def\@tempa{& &}
      \or \def\@tempa{&}
      \or\def\@tempa{}
\fi\@tempa
\if@eqnsw
\if@fewtab\@eqnnum\fi
\stepcounter{equation}\fi\global
\@eqnswtrue\global\@eqcnt\z@\global\@fewtabtrue\cr}

\def\draftcite#1{\ifnum\draftcontrol=1#1\else{}\fi}

\def\@lbibitem[#1]#2{\item{}\hskip -3cm \hbox to 2cm
{\hfil$\scriptstyle\draftcite{#2}$}\hskip
1cm[\@biblabel{#1}]\if@filesw
     {\def\protect##1{\string ##1\space}\immediate
      \write\@auxout{\string\bibcite{#2}{#1}}}\fi\ignorespaces}

\def\@bibitem#1{\item\hskip -3cm \hbox to 2cm
{\hfil $\scriptstyle\draftcite{#1}$}\hskip 1cm
\if@filesw \immediate\write\@auxout
       {\string\bibcite{#1}{\the\value{\@listctr}}}\fi\ignorespaces}

\font\tendl=msbm10  scaled \magstep1
\font\sevendl=msbm7 scaled \magstep1
\font\fivedl=msbm5 scaled \magstep1
\font\tengl=eufm10  scaled \magstep1
\font\sevengl=eufm7 scaled \magstep1
\font\fivegl=eufm5 scaled \magstep1

\newfam\dlfam  
\textfont\dlfam=\tendl \scriptfont\dlfam=\sevendl
\scriptscriptfont\dlfam=\fivedl
\newfam\glfam  
\textfont\glfam=\tengl \scriptfont\glfam=\sevengl
\scriptscriptfont\glfam=\fivegl

\def\draftdate{\number\month/\number\day/\number\year\ \ \ \hourmin }

\global\def\draftcontrol{0}
\catcode`\@=12
\def\tilde{\widetilde}
\def\hat{\widehat}
\documentstyle[12pt]{article}

\def\theequation{{\arabic{equation}}}

\newcommand{\be}{\begin{eqnarray}}
\newcommand{\en}{\end{eqnarray}\vs 0.5 cm}
\newcommand{\non}{\nonumber}
\newcommand{\no}{\noindent}
\newcommand{\vs}{\vskip}
\newcommand{\hs}{\hspace}

\newcommand{\un}{\underline}
\newcommand{\lsim}{{\smash{\mathop{<}\limits_{^\sim}}}}
\newcommand{\gsim}{{\smash{\mathop{>}\limits_{^\sim}}}}
\newcommand{\qq}{\begin{eqnarray}}

\newcommand{\da}{\partial}
\newcommand{\ee}{{\rm e}}

\newcommand{\qqq}{\end{eqnarray}}

\newcommand{\tr}{\hbox{tr}}

\newcommand{\CC}{{\cal C}}

\newcommand{\CF}{{\cal F}}

\newcommand{\CM}{{\cal M}}

\newcommand{\CO}{{\cal O}}

\newcommand{\CR}{{\cal R}}

\newcommand{\CV}{{\cal V}}

\newcommand{\la}{\lambda}
\newcommand{\La}{\Lambda}
\newcommand{\s}{\hspace{0.05cm}}
\newcommand{\m}{\hspace{0.025cm}}

\newcommand{\dd}{{{}^-\hs{-0.3cm}d}}

\begin{document}
\
\vskip 2.3cm
\no{\large{\bf{TURBULENCE UNDER A MAGNIFYING 
GLASS}}}\footnote{lectures given at the 1996 
Carg\`{e}se Summer Institute, July 22 - August 3}
\vskip 0.8cm
{}\hspace{2cm}Krzysztof Gaw\c{e}dzki
\vskip 0.3cm
{}\hspace{2cm}C.N.R.S., I.H.E.S.,

{}\hspace{2cm}Bures-sur-Yvette, 91440, France

\date{ }

\addtocounter{section}{1}
\vskip 1cm

\no{\large{\bf INTRODUCTION}}
\vskip 0.5cm

This is an introductory course on the
open problems of fully developed turbulence which present
a long standing challenge for theoretical and mathematical
physics. The plan of the course is as follows:
\vskip 0.2cm

\no {\bf{Lecture 1}}. \ Hydrodynamical equations.
Existence of solutions. Statistical description.
Kolmogorov scaling theory.
\vskip 0.2cm

\no {\bf{Lecture 2}}. \ Functional approach to turbulence,
similarities and differences with field theory.
\vskip 0.2cm

\no {\bf{Lecture 3}}. \ Passive scalar and breakdown
of the Kolmogorov theory.
\vskip 0.2cm

\no {\bf{Lecture 4}}. \ Inverse renormalization group.
\vskip 1cm

\no {\large\bf{LECTURE 1}}
\vskip 0.5cm

The hydrodynamical flows in gasses and liquids are
believed to be described in a variety of realistic
situations by incompressible Navier-Stokes (NS) equations
\qq
\da_t v\s+\s(v\cdot\nabla)\m v-\nu\Delta v\s=\s
{_1\over^\rho}\s(f-\nabla p)\s,\quad\ \ \ \nabla\cdot v\s=\s0\s
\label{NS}
\qqq
where the vector $v(t,x)$ describes the velocity
of the fluid at time $t$ and space point $x$,
positive constant $\nu$ is the viscosity, $\rho$ is the constant
density\footnote{we shall set it to $1$ below}, $f(t,x)$
is the (intensive) force and $p(t,x)$ is the pressure.
The above equations date back to the works of Navier (1823)
and Stokes (1843) and modify the even older Euler
equation (1755) by the addition of the dissipative
term \s$\nu\Delta v\m$. In most physical applications
the dimensionality of the space is $3$ or $2$, but
the equations make sense in general dimension $d$.
The Euler equation without the $\nu\Delta v$ term
is really the \s$F=ma\s$ (or rather $a={_1\over^m}\s F$\s)
\s equality for the volume element of the fluid.
The $\nu\Delta v$ term represents the friction forces.
By taking the $L^2$ scalar product of the NS equation
with $v$ and assuming that the flow velocity vanishes sufficiently
fast at large distances (or at the boundary),
one deduces the energy balance:
\qq
\da_t\s\m{_1\over^2}\int v^2\s=\s-\m\nu\int(\nabla v)^2
\s+\s\int f\cdot v\s.
\label{bal}
\qqq
The equation says that the rate of change
of fluid energy is equal to the energy injection $\int f\cdot v$
(the work of external forces per unit time) minus the
energy dissipation per unit time $\nu\int(\nabla v)^2$
due to the fluid friction.
\vskip 0.5cm

The Euler equation has a nice infinite-dimensional geometric
interpretation: it describes the geodesic flow on the
group of volume preserving diffeomorphisms\footnote{recall
that the Euler top describes the geodesic flow on the group
$SO(3)$}. We shall briefly sketch this argument.
Let, more generally, the space be an oriented $d$-dimensional
Riemannian manifold $(M,g)$. Denote by $\tau$ the Riemannian
volume: \s$\tau(x)=\sqrt{g(x)}\s dx^1\wedge\cdots\wedge dx^d\m$.
Let $Diff_\tau$ be the group of diffeomorphisms $\phi$
of $M$ preserving $\tau$: \s$\phi^*\tau=\tau$.
The space $Vect_\tau$ of divergenceless vector fields $v$
(i.e. the ones preserving $\tau$, \m${\cal L}_v\tau=0\s$)
\s may be considered as the Lie
algebra of $Diff_\tau$\footnote{for non-compact $M$
we shall assume that $\phi$ do not move points outside
a compact subset and that $v$ have compact support}.
Upon the identification of infinitesimal variations
of diffeomorphisms $\phi$ with the vector fields $v$
by \s$\delta\phi(x)=v(\phi(x))\m$, \s the
scalar product of vector fields
\qq
\Vert v\Vert^2\s=\s\int\limits_M g(v,v)\s\m\tau\s\equiv\s
\int\limits_M g_{ij}(x)\s v^i(x)\s v^j(x)\s\s\tau(x)
\non
\qqq
induces on $Diff_\tau$ a right-invariant Riemannian
metric. This metric in not left-invariant. Indeed,
\qq
dL_\phi\s dR_{\phi^{-1}}\s v\s=\s\phi_*v
\non
\qqq
where \s$dL_\phi\s$ and \s$dR_\phi$ stand for
the tangent maps to the left and right translations
by $\phi$ on $Diff_\tau$ and
\qq
\phi_*v(x)\s=\s d\phi\s v(\phi^{-1}(x))
\non
\qqq
is the pushforward of $v$ by $\phi$. But, in general,
\qq
\Vert\phi_*v\Vert^2\s\not=\s\Vert v\Vert^2
\non
\qqq
since the tangent map $d\phi$ does not preserve
the Riemannian length of vectors.
\vskip 0.3cm

The geodesic flows with respect to left-right-invariant
metrics on a group are, modulo time-independent left
and right translations, one-parameter subgroups.
This is not the case if the metric is only right-invariant.
The geodesic motions $\s t\mapsto\phi_t\s$
on $Diff_\tau$ corresponding to the right-invariant Riemannian
metric defined above extremize the action
\qq
S(\phi)\s=\s{_1\over^2}\int\Vert v(t)\Vert^2\s dt
\non
\qqq
where
\qq
\da_t\phi(t,x)\s=\s v(t,\phi(t,x))\s\equiv\s v(t,y)\s.
\non
\qqq
Explicitly\footnote{field theoriests will notice a similarity
of the following calculation to those in nonlinear sigma models},
\qq
&&\delta S(\phi)\s=\s{_1\over^2}\s\delta\int g_{ij}(y)
\s v^i(t,y)\s v^j(t,y)\s\s\tau(y)\s\s dt\cr
&&=\s{_1\over^2}\s\delta\int g_{ij}(\phi(t,x))\s\s\da_t\phi^i(t,x)\s\s
\da_t\phi^j(t,x)\s\s\tau(x)\s\s dt\cr
&&=\s{_1\over^2}\int\da_k g_{ij}(\phi(t,x))\s\s\delta\phi^k(t,x)\s\s
\da_t\phi^i(t,x)\s\s\da_t\phi^j(t,x)\s\s\tau(x)\s\s dt\cr
&&\hspace{4cm}+\s\int g_{ij}(\phi(t,x))\s\s
\da_t\phi^i(t,x)\s\s\da_t\delta
\phi^j(t,x)\s\s\tau(x)\s\s dt\cr
&&=\int\delta\phi^j(t,x)\bigg[{_1\over^2}\s\da_j g_{ik}(y)\s\m
v^i(t,y)\s\m v^k(t,y)\s-\s\da_k g_{ij}(y)\s\m v^i(t,y)\s\m
v^k(t,y)\cr
&&\hspace{3.4cm}-\s g_{ij}(y)\left(\m\da_t v^i(t,y)\m+\m v^k(t,y)\m
\da_k v^i(t,y)\right)\bigg]\tau(y)\s\s dt\cr
&&=\s-\int g_{ij}\s\m u^j\bigg[\da_t v^i\s+\s v^k\da_k v^i\s
+\s{_1\over^2}\m g^{in}(\da_k g_{ln}+\da_l g_{kn}-\da_n g_{lk})\s v^k
\s v^l\bigg]\tau\s\m dt\cr
&&=\s -\int g(\s u\s,\s\m\da_t+\nabla_vv\s)\s\m \tau\s\m dt
\non
\qqq
where \s$u^j(t,y)\equiv\delta\phi^j(t,x)\s$ is an arbitrary
divergenceless vector field.
\qq
(\nabla_wv)^i=w^k\da_k v^i+\{{_i\atop^{k\s l}}\}\s w^k\m v^l
\non
\qqq
denotes the covariant derivative of the vector field $v$ in the
direction of $w$ with the Levi-Civita symbols
\qq
\{{_i\atop^{k\s l}}\}\s=\s
{_1\over^2}\m g^{in}(\da_k g_{ln}+\da_l g_{kn}-\da_n g_{lk})\s.
\non
\qqq
Since a vector field on $M$ orthogonal to all vector fields
without divergence is a gradient, we obtain the (generalized)
Euler equation
\qq
\da_tv+\nabla_v v = -\nabla p
\non
\qqq
for the divergenceless vector fields $v$.
The NS term $\nu\Delta v$ still makes sense
in the geometric setup and may be added to the equation.
\vskip 0.5cm

The Euler and NS equations are examples of
nonlinear partial differential evolution equations.
After a century and a half study, they still pose major open
problems as far as the control of their solutions
is concerned. The most interesting questions
concern the short-distance (ultra-violet) behavior.
Suppose that we start from smooth initial data and
the force $f$ is smooth. For simplicity, let us assume
compact support of both (we may also consider the compact
space or the case with boundary conditions). It is known
that the smooth solutions of the so posed problems are unique
and exist for short time. Do they exist for all times?
The answer is positive in 2 dimensions for both Euler
and NS equations \cite{Wolib,Ger} but in 3
dimensions the answer is not known. It is expected to
be positive in the NS case. The opinions about
the Euler case (no blowup versus finite-time blowup
for special smooth initial conditions) are divided
and fluctuate in time.
\vskip 0.3cm

In an important 1933 paper on the NS equation Leray \cite{Leray} has
introduced the notion of weak solutions of the equation. A
vector field $v(t,x)$ locally in $L^2$ is a weak solution if
\qq
\int\left[\left(\da_tu^i+\nu\Delta u^i
+(\da_ju^i)v^j\right)v^i\s+\s u^if^i\right]\s=\s0\ \quad
{\rm and}\ \quad\int(\da_i\varphi)\m v^i\s=\s0
\non
\qqq
for any smooth divergenceless vector field $u$ and
any smooth function $\varphi$, both with compact supports.
Leray showed by compactness arguments existence of weak
global solutions of the 3-dimensional NS equations with additional
properties (e.g. with space derivatives locally square
integrable). The weak solutions are not unique (there are weak
solutions of the 2-dimensional Euler equation with compact
support \cite{Scheff,Shnir}).
\vskip 0.5cm

The NS equation is invariant under rescalings. Let
\qq
&&\tilde v(t,x)\s=\s\tau\m s^{-1}\s v(\tau t,sx)\s,\cr
&&\tilde f(t,x)\s=\s\tau^2s^{-1}\s f(\tau t,sx)\s,\cr
&&\tilde p(t,x)\s=\s\tau^2s^{-2}\s v(\tau t,sx)\s,\cr
&&\tilde\nu\s=\s\tau\m s^{-2}\s\nu\s.
\non
\qqq
If $v$ and $p$ solve the NS equation with viscosity $\nu$
and force $f$ then $\tilde v$ and $\tilde p$ give a solution
for viscosity $\tilde\nu$ and force $\tilde f$.
It is convenient to introduce the dimensionless version
of the (inverse) viscosity, the Reynolds number
\qq
R\s=\s{\delta_{_L}v\s L\over\nu}\s,
\non
\qqq
where $\delta_{_L}v$ is a characteristic velocity difference over
scale $L$ of the order of the size of the system. For example,
for the flow in a pipe of radius $L$,
we may take $\delta_{_L}v$ as velocity in the middle of the pipe
(minus the vanishing velocity on the wall of the pipe).
A basic phenomenological observation in hydrodynamics
is that for $R\lsim 1$ the flows are regular (laminar)
whereas $R\gg 1$ correspond to very irregular
(turbulent) flows with a rich set of scenarios occurring
for intermediate $R$. Note the scale-dependent character
of the Reynold number. Following \cite{Gallav},
define the running Reynolds number
\qq
R_r\s=\s\left({_1\over^{\vert\CO(r)\vert}}\int_{_{\CO(r)}}
\hs{-0.3cm}\vert\nabla v\vert^2\right)^{\hs{-0.1cm}1/2}
{r^2\over\nu}
\non
\qqq
where \s$\CO(r)\m=\m\{\m\s(s,y)\ \vert
\ \vert s-t\vert<{r^2\over\nu}\s,
\ \vert y-x\vert< r\s\}\s$. \s Note that we may rewrite
\qq
R_r\s=\s{\delta_rv\ r\over \nu}
\non
\qqq
where $\delta_rv$, the mean velocity difference on scale
$r$, is calculated by multiplying the mean square gradient
of $v$ over $\CO(r)$ by $r$. The best regularity result about
the weak solutions of the NS equation is due
to Caffarelli-Kohn-Nirenberg \cite{CKN} and says
that there exist $\epsilon>0$ such that if $R_r<\epsilon$
then the solution is smooth on the $\CO(\epsilon\m r)$
neighborhood of $(t,x)\m$ \cite{Gallav}. \s This
implies that for a weak solution, the Haussdorf dimension
of the set of singularities is $<\m 1$. \m Note the
spirit of the result in line with the phenomenological
characterization of laminar flows.
\vskip 0.5cm

For high Reynolds numbers it is reasonable to attempt
a statistical description of complicated turbulent flows.
In the theoretical approach, the statistics may be generated
by considering random initial data or/and random forcing.
$v(t,x)$ becomes then a random field. We shall be interested
in describing a stationary statistical state of the latter.
In such a state the mean overall energy of the fluid is
constant in time so that the energy balance equation
(\ref{bal}) implies that
\qq
\int\langle\m\nu\m(\nabla v)^2\rangle\s=\s
\int\langle\m f\cdot v\rangle\s,
\non
\qqq
where $\s\langle\ -\ \rangle$ denotes the ensemble
average, or, in a homogeneous state,
\qq
\bar\epsilon\s\equiv\s\langle\m\nu\m(\nabla v)^2\rangle
\s=\s\langle\m f\cdot v\rangle\s\equiv\s\bar\varphi
\label{bal1}
\qqq
where $\bar\epsilon$ denotes the mean dissipation rate
and $\bar\varphi$ the mean injection rate of energy,
both with dimension ${length^2\over time^3}$.
In the situation where the energy injection is a large
distance process (e.g. in the atmospheric turbulence)
one expects that for high $R$ a scale separation occurs
with the energy dissipation taking place on much smaller
distances. Pictorially, energy is transmitted to
the fluid by induction of large eddies on scale $L$
which subsequently break to smaller scale eddies
and so on passing energy to shorter and shorter
scales without substantial loss until the viscous
scale $\eta$ is reached where the friction dissipates
energy. Such a (Richardson \cite{Rich}) energy
cascade is then characterized by the integral scale
$L$, the viscous scale $\eta$ and the energy
dissipation rate $\bar\epsilon$. The scale ratio $L/\eta$
should grow with the Reynolds number. The interval of distance
scales $r$ satisfying $L\gg r\gg\eta$ is
called the inertial range. The cascade picture may be
formulated in more quantitative terms by introducing
the quantities \cite{Frisch}
\qq
\bar\epsilon_{\leq K}\s=\s\m\nu\int
\limits_{\vert k\vert\leq K}\bigg(\int\langle\m
\nabla v(0)\cdot\nabla v(x)\rangle\s\s\ee^{-i\s k\cdot x}
\s\s dx\bigg)\s\dd k
\non
\qqq
($\dd k\equiv{dk\over(2\pi)^{-d}}$) and
\qq
\bar\varphi_{\leq K}\s=\s\int
\limits_{\vert k\vert\leq K}\bigg(\int\langle\m
f(0)\cdot v(x)\rangle\s\s\ee^{-i\s k\cdot x}
\s\s dx\bigg)\s\dd k
\non
\qqq
interpreted as the mean dissipation and mean injection rate
in wavenumbers $k$ with $\vert k\vert\leq K$.
The injection of energy limited to distances $\gsim L$
means that, as a function  of $K$, $\s\bar\varphi_{\leq K}\s$
is close to $\bar\epsilon$ everywhere except for $K\lsim{1\over L}$
where it falls to zero with $K\to0$. Similarly, the cascade picture
should imply that $\bar\epsilon_{\leq K}$ is negligable
for $K\ll{1\over\eta}$ and then grows to $\bar\epsilon$.
The difference
\qq
\bar\varphi_{\leq K}-\bar\epsilon_{\leq K}\s\equiv\s
\bar\pi_{K}
\non
\qqq
has the interpretation of the energy flux out of the
wavenumbers $k$ with $\vert k\vert\leq K$. It should be
approximately constant and equal to $\bar\epsilon$
in the inertial range \s${1\over L}\ll K\ll{1\over\eta}\s$.
\vskip 0.5cm

In 1941, Kolmogorov \cite{K41} went one step further
by postulating that the equal-time correlators of velocity
differences over distances in the inertial range should be
universal functions of the latter and of the dissipation rate
$\bar\epsilon$. In particular for the structure functions
\qq
S_n(x)\s=\s\langle\m(v(x)-v(0))\cdot\hat x)^n\rangle
\non
\qqq
with $\hat x\equiv{x\over\vert x\vert}$, assuming also
isotropy of the turbulent state, one obtains
\qq
S_n(x)\s=\s C_n\s\bar\epsilon^{n/3}\m r^{n/3}
\label{K41}
\qqq
with $r\equiv\vert x\vert$ and universal constants $C_n$.
Indeed, the right hand side
is the only function of $\bar\epsilon$ and $r$
with the dimension $({length\over time})^n$. Kolmogorov
theory implies that the typical velocity difference
over distance $r$ behaves as ${\bar\epsilon}^{1/3}\m r^{1/3}$.
For the scale-dependent Reynolds number
one obtains then \s$R_r\sim{{\bar\epsilon}^{1/3}\m r^{4/3}
\over\nu}\m$. \s In particular, $R\cong R_L\sim
{{\bar\epsilon}^{1/3}\m L^{4/3}\over\nu}\s$ and \s$\CO(1)=R_\eta\sim
{{\bar\epsilon}^{1/3}\m \eta^{4/3}\over\nu}\s$
hence \s${\eta\over L}\sim R^{-3/4}\s$ and it decreases with $R$.
\vskip 0.4cm

For $n=3$ the relation (\ref{K41}) is essentially a rigorous result
with $C_3=-{4\over 5}$ in 3 dimensions. Let us sketch the latter
argument \cite{LaLi,Frisch}. Suppose for simplicity that
the force $f$ is a random Gaussian field with mean zero
and the covariance
\qq
\langle\m f^i(t,x)\s f^j(s,y)\rangle\s=\s\delta(t-s)\s\s
\CC^{ij}(x-y)
\label{rf}
\qqq
with $\da_i \CC^{ij}=0\s$. \s Inferring from the NS equation
that
\qq
v(t+\Delta t)\s=\s v(t)\s-\s(\m(v\cdot\nabla v)\m v
-\nu\Delta v+\nabla p\m)\vert_{_{t}}\Delta t\s
+\s\int_{_t}^{^{t+\Delta t}}
\hs{-0.5cm} f(s)\s\m ds\s+\s\CO(\Delta t)^2
\non
\qqq
the stationarity of $\s\langle\m v(t,x)\cdot v(t,y)\rangle\s$
implies that
\qq
&&
2\m\nu\s\langle\m\nabla v(x)\cdot\nabla v(y)\rangle
\s=\s-\m\nu\langle\m\Delta v(x)\cdot v(y)\rangle\s-\s
\nu\langle\m v(x)\cdot\Delta v(y)\rangle\cr\cr
&&=\s
-\m\langle\m(v(x)\cdot\nabla)\m v(x)\cdot v(y)\rangle
\s-\s\langle\m v(x)\cdot(v(y)\cdot\nabla)\m v(y)\m\rangle
\s+\s\tr\s\s\CC(x-y)\s
\non
\qqq
(the pressure does not contribute due to the transversality
of $v$). The first two terms on the right hand side
may be rewritten as
\qq
\m{_1\over^2}\s\nabla_x\cdot\langle\m
(v(x)-v(y))^2(v(x)-v(y))\m\rangle
\non
\qqq
so that we obtain the relation
\qq
\nu\s\langle\m\nabla v(x)\cdot\nabla v(0)\rangle
\s-\s{_1\over^4}\s\nabla_x\cdot\langle\m(v(x)-v(0))^2(v(x)-v(0))
\m\rangle\s=\s{_1\over^2}\s\tr\s\s\CC(x)\s.
\label{rel}
\qqq
Taking first the limit $x\to 0$ for positive $\nu$
and assuming that the presence of the latter smoothes
the behavior of \s$\langle\m(v(x)-v(0))^2(v(x)-v(0))\m\rangle\s$
so that the second term on the left hand side vanishes
when $x\to 0$, we obtain
\qq
\bar\epsilon\s=\s{_1\over^2}\s\tr\s\s\CC(0)
\non
\qqq
which is nothing else then the energy balance equation
(\ref{bal1}). On the other hand, taking first the $\nu\to 0$
limit of eq.\s\s(\ref{rel}) for $x\not=0$, we obtain
\qq
-\s{_1\over^4}\s\nabla_x\cdot\langle\m(v(x)-v(0))^2(v(x)-v(0))
\m\rangle\vert_{_{\nu=0}}\s=\s{_1\over^2}\s\tr\s\s\CC(x)\s.
\label{rel1}
\qqq
The assumption that the force acts only on distances $\gsim L$
means that for $r\ll L\m$, \s$\CC(x)\cong
\CC(0)\s$ so that eq.\s\s(\ref{rel1}) means that in the inertial
range
\qq
-\s{_1\over^4}\s\nabla_x\cdot\langle\m(v(x)-v(0))^2(v(x)-v(0))
\m\rangle\s=\s\bar\epsilon\s.
\non
\qqq
Together with isotropy, this implies
the relation \cite{Frisch}
\qq
\langle\m(v^i(x)-v^i(0))\m(v^j(x)-v^j(0))\m
(v^k(x)-v^k(0))\m\rangle
=-\m{_{4\s\bar\epsilon}\over^{d(d+2)}}
\s(\delta^{ij} x^k+\delta^{ik} x^j+\delta^{jk} x^i)\hs{0.7cm}
\non
\qqq
from which the $n=3$ case of eq.\s\s(\ref{K41}) follows
with \s$C_3=-\m{12\over d(d+2)}\m$.
\vskip0.5cm

The structure functions are measured, more or less
directly, in atmospheric or ocean flows, in water jets,
in aerodynamic tunnels or in subtle experiments
with helium gas inbetween rotating cylinders.
They are also accessible in numerical simulations.
One extracts then the scaling exponents assuming
the behavior
\qq
S_n(x)\s\sim\s r^{\zeta_n}\s.
\non
\qqq
$\zeta_3$ agrees well with the theoretical prediction
$\zeta_3=1$. Here are some other exponents obtained
from wind tunnel data \cite{Benzi}
\qq
\zeta_2=.70\s(.67)\m,\quad\hs{-0.06cm}\zeta_4=1.28\s(1.33)\m,
\quad\hs{-0.06cm}\zeta_5=1.53\s(1.67)\m,\quad\hs{-0.06cm}
\zeta_6=1.77\s(2)\m,\quad\hs{-0.06cm}\zeta_7=2.01\s(2.33)
\non
\qqq
with the Kolmogorov values in the parenthesis for comparison.
The discrepancy is quite pronounced (its direction is determined
by the H\"{o}lder inequality). One of the main open problems
in the theory of fully developed turbulence is to explain,
starting from the first principles (i.e.\s\s from the NS equation),
the breakdown of the Kolmogorov theory leading to the anomalous
structure-function exponents which indicate that the distribution
of $v(t)$ in the inertial range is far from Gaussian. The intuitive
idea that only a part of fluid modes (temporal or/and spacial)
participates in the turbulent cascade ("intermittency") has
led to multiple models of the cascade of the (multi)fractal
type \cite{Frisch}. Such models are not based on the NS equation
and allow to obtain essentially arbitrary spectra of
(multifractal \cite{DupLud}) exponents
hence they do not really explain the breakdown of the normal
scaling in realistic flows.
\vskip 0.4cm

For the "energy spectrum"
\qq
e(K)\s\equiv\s{_1\over^2}\s{_d\over^{dK}}
\int\limits_{\vert k\vert\leq K}\hs{-0.2cm}
\bigg(\int\langle\m v(0)\cdot v(x)\rangle\s\s\ee^{-i\s k\cdot x}
\s\s dx\bigg)\s\dd k\s,
\non
\qqq
the Kolmogorov theory predicts
\qq
e(K)\sim\s{\bar\epsilon}^{2/3}\s K^{-5/3}
\non
\qqq
for \s${1\over L}\ll K\ll{1\over\eta}\m$.
The experimental data seem to confirm this behavior (with
the possibility of a slight discrepancy consistent with the
above value of $\zeta_2$). Deep in the dissipative regime
($K\gg{1\over^\eta}$), \s$e(K)$ falls off much
faster than in the inertial regime.
\vskip 1cm

\no {\large\bf{LECTURE 2}}
\addtocounter{section}{1}
\vskip 0.5cm

The NS equation with random force is an example of a stochastic
evolution equation of the form
\qq
\da_t\Phi\s=\s G(\Phi)\s+\s f
\label{ee}
\qqq
with $\Phi$ a function of time, $G(\Phi)$ a functional
local in time and \s$f$ a stationary
Gaussian process with mean zero and covariance
\qq
\langle\s f(t)\s\m f(s)\s\rangle\s=\s C(t-s)\s.
\non
\qqq
Upon solving the equation for given $f$ with initial data
$\Phi(t_0)=\Phi_0$, one obtains a stochastic process
\s$t\mapsto\Phi(t)$. The limit $t_0\to-\infty$, if exists,
should then allow to pass to the stationary regime without
dependence on the initial data. There is a simple way to write
the expectation values of a functional $\CF$ of $\Phi$ in terms
of a formal functional integral:
\qq
\langle\s\CF(\Phi)\s\rangle\s=\s\int\CF(\Phi)\s\s\delta(\da_t\Phi
-G(\Phi)-f)\s\s\det(\da_t-{_{\delta G}\over^{\delta\Phi}})
\s\s D\Phi\s\s d\mu_{_C}(f)
\label{ee1}
\qqq
where $d\mu_{_C}$ denotes the Gaussian measure with covariance
$C$. Indeed, the integration over $\Phi$ imposes
the dependence (\ref{ee}) between $\Phi$ and $f$ and
the $f$ integral calculates the Gaussian expectation of $\CF(\Phi)$
viewed as a functional of $f$. The determinant in (\ref{ee1})
is calculable (formally).
\qq
&&\det(\da_t-{_{\delta G}\over^{\delta\Phi}})\s=\s
\det(\da_t)\s\s\det(\m 1\m-\m\da_t^{-1}{_{\delta G}
\over^{\delta\Phi}})\cr
&&=\s\det(\da_t)\ \ee^{\s{\rm tr}\s\ln(1-\da_t^{-1}
{{\delta G}\over{\delta\Phi}})}\s=\s\det(\da_t)
\ \ee^{\m-\sum\limits_{n=1}^\infty{\rm tr}\s(\da_t^{-1}
{{\delta G}\over{\delta\Phi}})^n}\s.
\non
\qqq
With the choice $\da_t^{-1}(t_1,t_2)=\theta(t_1-t_2)$,
appropriate for solutions of the initial value problems,
we have, setting\footnote{in simplest situations, this
choice cancels in final expressions, in more complicated ones
it is related to a choice of the (ordering) convention for
the stochastic integrals} $\theta(0)={1\over2}$,
\qq
\tr\s\s(\da_t^{-1}{_{\delta G}\over^{\delta\Phi}})^n\ =\
\cases{\hbox to 3cm{0\hfill}{\rm for}\quad n>1\s,\cr
\hbox to 3cm{${1\over 2}\int{\rm tr}\s{{\delta G}
\over{\delta\Phi}}\s dt$\hfill}{\rm for}\quad n=1\s,}
\non
\qqq
where on the right hand side, \s${{\delta G}\over{\delta\Phi}}\s$
is calculated at fixed time. Inserting the value of the determinant
into eq.\s\s(\ref{ee1}) and rewriting the $\delta$-function
as an oscillatory integral, we obtain
\qq
\langle\s\CF(\Phi)\s\rangle\s=\s\int\CF(\Phi)\ \ee^{\m-\m
i\s(\m R\s,\s\m\da_t\Phi-G(\Phi)-f\s)\s-\s{1\over 2}\int\tr\s
{{\delta G}\over{\delta\Phi}}\s dt}\ DR\s\s D\Phi\s\s d\mu_{_C}(f)
\ \bigg/\ {\rm norm.}\s,
\non
\qqq
where $(\s\cdot\s,\s\cdot\s)$ stands for the $L^2$ scalar product,
or, after integration over $f$,
\qq
\langle\s\CF(\Phi)\s\rangle\s=\s\int\CF(\Phi)\ \ee^{\s-\m S(R,\Phi)}
\ DR\s\s D\Phi\ \bigg/\ {\rm norm.}
\label{MSR}
\qqq
where
\qq
S(R,\Phi)\s=\s i\s(\s R\s,\s\m\da_t\Phi-G(\Phi)\s)\s
+\s{_1\over^2}\int\tr\s
{_{\delta G}\over^{\delta\Phi}}\s dt\s+\s{_1\over^2}
\s(R,\m C R)\s.
\label{MSR1}
\qqq
The above functional integral representation is known in the
physical literature as the Martin-Siggia-Rose (MSR) formalism
\cite{MSR}. The field $R$ is called the response field
(its correlations measure the reaction of the system to small
deterministic variations of the force).
It may be integrated out in (\ref{MSR}) leading to a Fokker-Planck
type of functional integral but it will be more convenient
to keep it in the functional representation.
\vskip 0.5cm

As a simple but instructive example, let us consider
a linear stochastic equation describing forced diffusion:
\qq
\da_t{T}\s=\s D\m\Delta{T}\s+\s f
\label{fd}
\qqq
with the diffusion constant $D$ and the forcing covariance
\qq
\langle\s f(t,x)\s\m f(s,y)\s\rangle\s=\s\delta(t-s)\s\m\CC(x-y)
\label{for}
\qqq
where $\CC$ is a smooth, fast decaying, positive-definite function.
Given the initial data ${T}(t_0)$, we may solve eq.\s\s(\ref{fd}):
\qq
{T}(t,x)\s=\s\int\ee^{\m(t-t_0)D\Delta}(x,y)\s\s{T}(t_0,y)\s\s dy
\s+\s\int\limits_{t_0}^t ds\int\ee^{\m(t-s)D\Delta}(x,y)\s\s f(s,y)
\s\s dy
\non
\qqq
obtaining a Gaussian stochastic process with mean
\qq
\ee^{\m(t-t_0)D\Delta}\s{T}(t_0)
\non
\qqq
and covariance
\qq
&&\langle\s({T}(t_1,x_1)-\langle{T}(t_1,x_1)\rangle)\s({T}(t_2,x_2)
-\langle{T}(t_2,x_2)\rangle)\s\rangle\cr
&&\hs{3cm}=\s
\int\limits_{t_0}^{{\rm min}(t_1,t_2)}\left(\ee^{\m(t_1-s)D\Delta}
\s\CC\s\ee^{\m(t_2-s)D\Delta}\right)(x_1,\m x_2)\s\s ds\s.
\non
\qqq
The limit $t_0\to-\infty$ results in a stationary Gaussian
process with mean zero and covariance
\qq
\langle\s{T}(t_1,x_1)\s\m{T}(t_2,x_2)\s\rangle\s=\s
\int\ee^{\m\vert t_1-t_2\vert\m D\m k^2\s-\s i\m k\cdot(x_1-x_2)}
\s\s{_{\hat\CC(k)}\over^{2Dk^2}}\s\s\dd k\s.
\label{co}
\qqq
Note that for $t_1=t_2$ the right hand side is
a propagator of the scalar massless
free field with UV cutoff $\hat\CC(k)$.
\vskip 0.3cm

In the MSR formalism, the stochastic equation (\ref{fd})
leads to the quadratic action
\qq
S(R,{T})= i\hs{-0.1cm}\int\hs{-0.13cm}R(t,x)\m
(\da_t{T}-D\Delta{T})(t,x)\s
dt\s dx\m+\m{_1\over^2}\hs{-0.1cm}\int\hs{-0.13cm}
R(t,x)\m\CC(x-y)\m R(t,y)\s dt
\s dx\s dy\hs{0.6cm}
\non
\qqq
which gives rise to the propagators
\qq
\langle\s R\s\s R\s\rangle\hs{2.4cm}&=&0\s,\cr\cr
\langle\m\s {T}(t_1,x_1)\s\s R(t_2,x_2)\s\rangle&=&
i\m(\da_t+D\Delta)^{-1}(t_1,x_1;\m t_2,x_2)\cr
&=&i\m\theta(t_1-t_2)\s\s\ee^{\m(t_1-t_2)D\Delta}(x_1,x_2)
\label{co1}
\qqq
and $\langle{T}\m{T}\rangle$ as in eq.\s\s(\ref{co}).
\vskip 0.3cm

There are two interesting limiting regimes of forced diffusion:
the one of long times and long distances and that of
short times and short distances. To study the first one, let
us introduce the rescaled fields
\qq
{T}_\la(t,x)\s=\s\la^{(d-2)/2}\s{T}(\la t^2,\m\la x)\s,
\quad\quad R_\la(t,x)\s=\s\la^{(d+2)/2}\s R(\la^2 t,\m \la x)
\label{resc1}
\qqq
for which
\qq
&&\langle\s{T}_\la(t_1,x_1)\s\m{T}_\la(t_2,x_2)\s\rangle\s=\s
\la^{d-2}\int\ee^{\m\vert t_1-t_2\vert\m\la^2 D\m k^2\s
-\s i\m k\cdot(x_1-x_2)\m\la}
\s\s{_{\hat\CC(k)}\over^{2Dk^2}}\s\s\dd k\cr
&&=\s
\int\ee^{\m\vert t_1-t_2\vert\m D\m k^2\s
-\s i\m k\cdot(x_1-x_2)}
\s\s{_{\hat\CC(k/\la)}\over^{2Dk^2}}\s\s\dd k\s
\ \ {\smash{\mathop{\longrightarrow}\limits_{\la\to\infty}}}\ \s\
{_{\hat\CC(0)}\over^{2D}}\int\ee^{\m-\vert t_1-t_2\vert\m D\m k^2\s
+\s i\m k\cdot(x_1-x_2)}\s\s{_1\over^{k^2}}\s\s\dd k\s,
\hs{1.1cm}\label{lr1}
\\\cr
&&\langle\s{T}_\la\s\m R_\la\s\rangle\s=\s\langle\s{T}\s\s R
\s\rangle\s.\nonumber
\qqq
The limiting $\s\langle{T}_\la{T}_\la\rangle\s$ covariance
coincides with the one of the Langevin dynamics (of type A
\cite{HalHo}) for the scalar massless field.
Recall that the Langevin dynamics
is given by the stochastic evolution equation
\qq
\da_t\Phi\s=\s-\m{_1\over^2}\s
\Gamma\s{_{\delta\CV({\Phi})}\over^{\delta{\Phi}}}
\s+\s f
\non
\qqq
with the delta-correlated noise
\qq
\langle\s f(t,x)\s\m f(s,y)\s\rangle\s=\s\Gamma\s\m\delta(t-s)\s
\m\delta(x-y)\s.
\non
\qqq
For \s$\CV(\Phi)=\int V(\Phi(t))\s dt\m$, \s it
describes convergence to the equilibrium (equal time)
probability measure \s\s$\sim\s\ee^{\m-V(\Phi)}\s D\Phi\m.\s$
In our case
\qq
V({T})\s=\s {_D\over^{\hat\CC(0)}}\int(\nabla{T})^2\s,\quad
\quad\ \ \Gamma\s=\s\hat\CC(0)\s.
\non
\qqq
The limiting behavior (\ref{lr1}) expresses the simple fact that
at large distances the noise covariance $\CC(x-y)$ looks like
$(\int\CC)\s\m\delta(x-y)\m.$
\vskip 0.3cm

The study of the long-time, long-distance asymptotics is
a typical field-theoretic problem where we are interested
in the behavior of the system at distances much longer
than the UV cutoff scale. Studying the opposite regime
of distances much shorter than the cutoff scale may seem
without interest. This is not so. In problems related to
turbulence, function $\hat\CC(k)$ describes the spectrum
of forcing rather than the momentum cutoff and we are
interested in the regime of short times and short distances,
more exactly in distance scales much shorter than the
scale of energy injection interpreted in the field-theoretic
context as the UV cutoff scale. Hence {\bf field theory and
turbulence are concerned with opposite limiting regimes}.
In order to examine the short-time, short-distance
asymptotics of forced diffusion, we rescale the fields
differently introducing
\qq
{T}^\la(t,x)\s=\s\la\s{T}(t/\la^2,\m x/\la)\s,\quad\quad
R^\la(t,x)\s=\s\la^{-d-1}\s R(t/\la^2,\m x/\la)\s.
\label{resc2}
\qqq
and obtaining
\qq
&&\langle\s{T}^\la(t_1,x_1)\s\m{T}^\la(t_2,x_2)\s\rangle\s=\s
\la^2\int\ee^{\m\vert t_1-t_2\vert\m\la^{-2} D\m k^2\s
-\s i\m k\cdot(x_1-x_2)\m\la^{-1}}
\s\s{_{\hat\CC(k)}\over^{2Dk^2}}\s\s\dd k\cr
&&=\s\la^2\int{_{\hat\CC(k)}\over^{2Dk^2}}\s\s\dd k\ -\
{_1\over^2}\s\CC(0)\left(\vert t_1-t_2\vert\s+\s{_1\over^{Dd}}
\s\vert x_1-x_2\vert^2\right)\s+\s\CO(\la^2)\s,\label{lr2}\\\cr
&&\langle\s{T}^\la\s\m R^\la\s\rangle\s=\s\langle\s{T}\s\s R
\s\rangle\nonumber
\qqq
for large $\la$. Hence the $\langle{T}^\la\m{T}^\la\rangle$
covariance reaches a scaling form modulo a divergent constant.
In other words, the differences ${T}(t,x)-{T}(t,y)$ exhibit
a scaling behavior at short times and short distances.
The presence of the blowing up constant mode which may
be eliminated by considering field differences is typical
for the turbulence related problems.
\vskip 0.5cm

Returning to the NS equation (\ref{NS}) with random
Gaussian force (\ref{rf}) and applying
the MSR formalism, we obtain the MSR action
\qq
S(R,v)\s=\s i\int R(t,x)\cdot(\da_t v+v\cdot\nabla v-\nu\Delta v)(t,x)
\m\s dt\s dx\cr
+\s{_1\over^2}\int R(t,x)\cdot\CC(x-y)\m R(t,y)\s\m dt\s dx\s dy
\non
\qqq
with vector fields $R$ and $v$ satisfying $\nabla\cdot R=0$
and $\nabla\cdot v=0$. One could then set up a perturbative
scheme by separating
\qq
S(R,v)\s=\s S_0(R,v)\s+\s S_1(R,v)
\non
\qqq
with
\qq
&&S_0(R,v)=
i\hs{-0.1cm}\int\hs{-0.1cm}R(t,x)\cdot(\da_t v-\nu\Delta v)(t,x)
\s dt\s dx\m+\m{_1\over^2}\hs{-0.1cm}\int\hs{-0.1cm}
R(t,x)\cdot\CC(x-y)\m R(t,y)\s dt\s dx\s dy\m,\cr
&&S_1(R,v)= i\hs{-0.1cm}\int\hs{-0.1cm}R\s P_{_{\rm tr}}
(v\cdot\nabla v)\s\m dt\s dx\m,
\non
\qqq
(\m$P_{_{\rm tr}}\s$ stands for the projection on
the transverse vector fields)
and developing $\s\ee^{-S_1}\s$ in the power series,
with \s$S_0\s$ giving rise to the propagators
\qq
\langle R\s v\rangle_{_0}\quad\quad\Large{-\hs{-0.25cm}-
\hs{-0.12cm}\cdot\hs{-0.1cm}\cdot\hs{-0.1cm}\cdot\hs{-0.02cm}\cdot}
\quad,\quad\quad\quad
\langle v\s v\rangle_{_0}\quad\quad\Large{\cdot\hs{-0.1cm}
\cdot\hs{-0.1cm}\cdot\hs{-0.1cm}\cdot\hs{-0.1cm}\cdot\hs{-0.11cm}
\cdot\hs{-0.1cm}\cdot\hs{-0.0cm}\cdot}
\non
\qqq
and $S_1\s$ to the vertex
\qq
\Large{-\hs{-0.25cm}-\hs{-0.13cm}\cdot\hs{-0.13cm}:\hs{-0.15cm}
{_.\atop{^{^\cdot}}}\hs{-0.1cm}{_\cdot\atop^\cdot}}
\non
\qqq
and with the simplest Feynman diagrams
\qq
\Large{-\hs{-0.25cm}-\hs{-0.13cm}\cdot\hs{-0.01cm}
\smash{\mathop{_{^.}}\limits^{\cdot}}\hs{-0.01cm}
\smash{\mathop{_{^.}}\limits^{^.}}\hs{-0.01cm}
\smash{\mathop{_{^.}}\limits^{^\cdot}}\hs{-0.01cm}
\smash{\mathop{_{^.}}\limits^{^\cdot}}\hs{-0.03cm}
\smash{\mathop{_{^{_{-}}}}\limits^{\cdot_{\hs{0.03cm}\cdot}}}
\hs{-0.17cm}\smash{\mathop{_{^{_{-}}}}\limits^{\hs{0.08cm}_.}}
\hs{-0.17cm}\cdot\hs{-0.1cm}\cdot\hs{-0.1cm}\cdot\hs{-0.1cm}\cdot}
\quad,\quad\quad\quad
\Large{-\hs{-0.25cm}-\hs{-0.13cm}\cdot\hs{-0.13cm}:\hs{-0.15cm}
{_.\atop{^{^\cdot}}}\hs{-0.08cm}{_{_{^\cdot}}\atop^{^{_\cdot}}}
\hs{-0.1cm}{_{_{^\cdot}}\atop^{^{_\cdot}}}\hs{-0.12cm}
{\hs{0.02cm}_.\atop{^{^\cdot}}}\hs{-0.18cm}:\hs{-0.13cm}\cdot\hs{-0.13cm}-
\hs{-0.25cm}-}\quad.
\non
\qqq
The perturbative expansion is plagued by the UV and IR
divergences when $\nu\to 0$ and $\CC\to{\rm const}.$
Attempts were made to improve the situation by
applying various resummation schemes, inspired
by the field theoretic techniques (Schwinger-Dyson
equations, renormalization group, etc.) but it seems fair
to say that they were not very successful.
\vskip 0.4cm

Despite a formal similarity with the field theory formulation
using functional integrals, the problem of the NS turbulence
differs radically from the (dynamical formulation) of
field theory in at least two crucial aspects:
\vskip 0.2cm

I. \ the nonlinear term \s$v\cdot\nabla v\s$ in the NS equation is
not of the gradient type \s${\delta\CV(v)\over\delta v}\s$
unlike the nonlinearities in the Langevin dynamics
for field theory models,
\vskip 0.2cm

II.\s\ the noise (force) correlation in the NS turbulence
should be close to a delta-function in the Fourier space
rather than in the position space as for the Langevin
dynamics.
\vskip 0.2cm

\noindent Consequently, the stationary statistical state of the
NS fluid is not of an equilibrium type unlike the Gibbs states
corresponding to the stationary states of the Langevin
dynamics for (euclidean) field theories. We have seen
on the example of linear forced diffusion that
the 2$^{\s{\rm nd}}$ difference is already enough to make
the problem very different from the field theoretic one.
Let us profit from the occasion to mention that
the study of the nonlinear version of forced
diffusion:
\qq
\da_t{T}\s=\s D\m\Delta{T}\s-\s{_1\over^{2\hat\CC(0)}}
\s{_{\delta\CV({T})}\over^{\delta{T}}}\s+\s f
\label{fdn}
\qqq
with random Gaussian force as in (\ref{for})
is a problem of its own interest. The long-time
long-distance asymptotics of the correlations
is then a dynamical version of an interacting
field theory problem (e.g. for \s$\CV({T})\sim\int{T}^4\m$)
and may be studied by the renormalization group
techniques. On the other hand, the short-time, short-distance
asymptotics e.g. for \s$\CV({T})\sim\int(\nabla{T})^4\m$
is a completely open problem\footnote{
we consider functionals of $\nabla{T}$ to avoid coupling
to the constant mode unstable at short distances already in
the linear case}. The inverse renormalization group
which will be the subject of the last lecture may provide
a tool for the latter type of questions.
\vskip 1cm

\no {\large\bf{LECTURE 3}}
\addtocounter{section}{1}
\vskip 0.5cm

In view of the reputed difficulty of the NS problem,
it would be instructive to consider simpler models of the
stochastic evolution equation (\ref{ee})
randomly forced at long distances and with $G(\Phi)$
not of the gradient type.
Such a model is provided by an equation, with a relatively
long history \cite{Obuh,Batch,Corrs}, describing
the passive advection of a scalar quantity $T(t,x)$
(temperature) by a random velocity field
\qq
\da_t T\s=\s -\m v\cdot\nabla T\s+\s\kappa\Delta T\s+\s f\s.
\label{ps}
\qqq
The positive coefficient $\kappa$ is called the molecular
diffusivity. For a divergence-free $v$, operator $-v\cdot\nabla$
is skew-symmetric (conserving the energy ${_1\over^2}\int T^2$)
and the $-v\cdot\nabla T$ term
is not of the gradient type, unlike the $\kappa\Delta T$
one corresponding to a negative (energy dissipating)
symmetric operator.
Ideally, the velocity field $v(t,x)$ should describe a turbulent
NS flow but, to simplify radically the problem, one replaces
it by a Gaussian random field. In \cite{Kr68} Kraichnan
noticed that in the case when $v$ is decorrelated in time
the problem becomes exactly soluble in the sense that one may write
a closed system of differential equations for the correlation
functions of $T$. In recent years Kraichnan's model
of passive scalar has attracted a lot of attention, see e.g.
\cite{Kr94,Proc1,Falk1,ShrS,GK}, as a prototype of a turbulent
system in which one may study analytically
the breakdown of a Kolmogorov-type scaling.
\vskip 0.3cm

We shall describe some results of the approach developed
in \cite{GK0,GK,BGK}. The Gaussian velocity field $v$ will
be taken of mean zero and covariance
\qq
\langle\s v^i(t,x)\s\m v^j(s,y)\s\rangle\s=\s
\delta(s-t)\s D^{ij}(x-y)
\label{covv}
\qqq
with $\da_i D^{ij}(x)=0$ and
\qq
D^{ij}(x)\s=\s2\m D_0\s\delta^{ij}\s-\s2\m d^{ij}(x)
\label{Dd}
\qqq
where $d^{ij}(x)\s\sim\s r^\xi$ for small $r\equiv|x|$,
\m$0<\xi<2$. Note the implied growth of the $2^{\s\rm nd}$
order velocity structure function with the distance,
as in the turbulent flows. $\xi={4\over 3}$
corresponds to the Kolmogorov scaling
dimension\footnote{as we shall see in the
next lecture, the dimension of $\delta(t-s)$ is $\xi-2$}
of $v$ equal to ${1\over 3}$. More concretely, we may pose
\qq
D^{ij}(x)\s\sim\s\int{{\ee^{i\m k\cdot x}}
\over{(k^2+m^2)^{(d+\xi)/2}}}\s\s\left(\delta^{ij}
-{_{k^ik^j}\over^{k^2}}\right)\m\dd k\s,
\label{covv1}
\qqq
with a small IR regulator $m$, which leads to $D_0\sim m^{-\xi}$ and
\qq
d^{ij}(x)\equiv d^{ij}_m(x)\s\cong
\s{_D\over^{d-1}}\left((d-1+\xi)\delta^{ij}
-\xi{_{x^ix^j}\over^{r^2}}\right)r^\xi\equiv d_0^{ij}(x)
\label{scar}
\qqq
for small $r$ and some constant $D$. Unlike $D_0$ which
diverges when $m\to0$, $d_m^{ij}(x)$
possesses the $m\to0$ limit equal to $d^{ij}_0(x)$
and scaling exactly with power $\xi$.
\vskip 0.2cm

The source $f$ will be assumed independent of $v$, Gaussian,
with mean zero and with covariance
\qq
\langle\s f(t,x)\s\m f(s,y)\s\rangle\s=\s
\delta(t-s)\s\m\CC({_{x-y}\over^L})
\non
\qqq
for some positive definite, rotationally invariant
test function $\CC$.
\vskip 0.3cm

The passive scalar model as set up above exhibits
scale separation with the energy cascade in the
inertial range of distances \s$L\gg r\gg\eta
=\CO(({\kappa\over D})^{1/\xi})\m L\m$. The
even structure functions of the scalar\footnote{the odd ones
vanish} \s$S_{2n}(r)\m=\m
\langle\m(T(t,x)-T(t,0))^{2n}\m\rangle\s$ show
exponential scaling in the inertial range. At least for small
$\xi$,
\qq
S_{2n}(r)\ \sim\ r^{\zeta_{2n}}
\label{strf}
\qqq
with
\qq
\zeta_{2n}\s=\s n(2-\xi)\s-\s{_{2n(n-1)}\over^{d+2}}\s\xi
\s+\s\CO(\xi^2)\s.
\label{zeta}
\qqq
In particular, the scaling of the 4$^{\s\rm th}$ and higher
structure functions is anomalous: the exponents $\zeta_{2n}$
deviate from the ones of the Kolmogorov(-Corrsin)
scaling theory giving $\zeta_{2n}=n\zeta_2$. Similar
results were obtained in \cite{Falk1,Falk2} for
large space dimension $d$.
\vskip 0.3cm

The above conclusions are based on an analysis of
the differential equations satisfied by the equal-time correlation
functions of the scalar $T$. These equations may be obtained
in many ways, for example with help of the MSR formalism.
Let us first consider the example of the 2-point function.
We have
\qq
\langle\s T(t,x_1)\s\m T(t,x_2)\s\rangle\s=\s
\int\left(\int T(t,x_1)\s T(t,x_2)\s\s\ee^{-S(R,T,v)}
\s\s DR\s DT\s\bigg/\s{\rm norm}.\right)d\mu_{_D}(v)
\hs{0.4cm}
\non
\qqq
where
\qq
S(R,T,v)= i\hs{-0.1cm}\int\hs{-0.1cm}R\s(\da_tT
+v\cdot\nabla T-\kappa\Delta T)
\m dt\m dx\m+\m{_1\over^2}\hs{-0.15cm}\int\hs{-0.15cm}
R(t,x)\m\CC({_{x-y}\over^L})\m R(t,y)\m dt\m dx\m dy\hs{0.8cm}
\non
\qqq
and $d\mu_{_D}(v)$ is the Gaussian measure corresponding
to the covariance (\ref{covv}). The normalization of the
Gaussian $R,\m T$ functional integral may be argued
to be $v$-independent. Performing this integral, one obtains
\qq
&&\langle\s T(t,x_1)\s\m T(t,x_2)\s\rangle\cr\cr
&&\hs{-0.4cm}=\s
\int\left((\da_t+v\cdot\nabla-\kappa\Delta)^{-1}\CC\m
(-\da_t+v\cdot\nabla-\kappa\Delta)^{-1}\right)
(t,x_1;\m t,x_2)\ d\mu_{_D}(v)\s.\hs{0.5cm}
\non
\qqq
Expanding into the Neuman series
\qq
(\pm\da_t+v\cdot\nabla-\kappa\Delta)^{-1}
\s=\s\sum\limits_{m=0}^\infty(\pm\da_t-\kappa\Delta)^{-1}
\left(- v\cdot\nabla\s(\pm\da_t-\kappa\Delta)^{-1}\right)^m\m,
\label{Neu}
\qqq
we may easily compute the $d\mu_{_D}(v)$ expectation
using the Wick theorem.
The resulting sum may be identified\footnote{the time-decorrelation
of the velocities is crucial at this point} with the Neuman series
for $\CM_2^{-1}\m\CC$ where $\CM_2$ is an operator
acting on the functions of $x_1,x_2$,
\qq
\CM_2&=&-\m\kappa(\Delta_{x_1}+\Delta_{x_2})
\s-\s{_1\over^{2\delta(0)}}\left(\langle(v(t,x_1)
\cdot\nabla_{x_1})^2\rangle
+\langle(v(t,x_2)\cdot\nabla_{x_2})^2\rangle\right)\cr
&&\hs{5cm}-\s{_1\over^{\delta(0)}}\s\langle
v(t,x_1)\cdot\nabla_{x_1}\s
v(t,x_2)\cdot\nabla_{x_2}\rangle\cr\cr
&=&-(\kappa+D_0)(\Delta_{x_1}+\Delta_{x_2})
\s-\s2\m(D_0\delta^{ij}-d^{ij}(x_1-x_2))
\s\da_{x_1^i}\da_{x_2^j}\s\m.
\non
\qqq
The Neuman series develops $\CM_2^{-1}$ in powers
of the $v$-expectations. The "tadpole" term
$\langle(v(t,x_i)\cdot\nabla_{x_i})^2\rangle\s$
originates from the Wick contraction of 2 velocities in the
Neuman series of the same \s$(\pm\da_t+v\cdot\nabla-\kappa
\Delta)^{-1}\s$ operator
whereas \s$\langle v(t,x_1)\cdot\nabla_{x_1}\s
v(t,x_2)\cdot\nabla_{x_2}\rangle\s$ comes from
the mixed contractions.
\vskip 0.3cm

Higher point correlation functions can be treated the same
way:
\qq
&&\hs{0.4cm}\CF_{2n}(x_1,\dots,x_{2n})\s\ \equiv\ \s
\langle\s T(t,x_1)\s\cdots\s T(t,x_{2n})\s\rangle\cr
&&=\hs{-0.42cm}\sum\limits_{{{\rm pairings}
\atop(\{n_j,m_j\})}}\hs{-0.32cm}
\int\hs{-0.1cm}\prod\limits_{j=1}^n\hs{-0.1cm}
\left((\da_t+v\cdot\nabla-\kappa\Delta)^{-1}\CC\m
(-\da_t+v\cdot\nabla-\kappa\Delta)^{-1}\right)\hs{-0.06cm}
(t,x_{n_j};\m t,x_{m_j})\ d\mu_{_D}(v)\hs{1.15cm}
\non
\qqq
and using eq.\s\s(\ref{Neu}) and the Wick theorem, it
is easy to obtain the inductive relation
\qq
\CF_{2n}(x_1,\dots,x_{2n})\s=\s\CM_{2n}^{\s-1}
\sum\limits_{1\leq n_1<n_2\leq{2n}}\CC({_{x_{n_1}-x_{n_2}}
\over^{L}})\ \CF_{2n-2}(x_1,\smash{\mathop{\dots\dots\dots}
\limits_{\hat{n_1}\quad\hat{n_2}}},x_{2n})
\label{2npt}
\qqq
(in a slightly abusive notation) with
\qq
\CM_{n}\ =\ -(\kappa+D_0)\sum\limits_{m=1}^n
\Delta_{x_m}\s-\ 2\hs{-0.4cm}\sum\limits_{1\leq m_1<m_2\leq n}
\hs{-0.2cm}(D_0\delta^{ij}-d^{ij}(x_{m_1}-x_{m_2}))
\s\da_{x_{m_1}^i}\da_{x_{m_2}^j}\s.\hs{0.5cm}
\non
\qqq
For $\kappa>0$, \s$\CM_n$ are positive 2$^{\s\rm nd}$
order elliptic operators. For $\kappa=0$ they become
singular elliptic (their principal symbol
looses positive definiteness at coinciding
points $x_{m_1}=x_{m_2}$). Note that
in the action on translationally invariant functions of
$x_1,\dots,x_n$,
\qq
\CM_n\ =\ -\m\kappa\sum\limits_{m=1}^n\Delta_{x_m}\s+\ 2\hs{-0.4cm}
\sum\limits_{1\leq m_1<m_2\leq n}\hs{-0.2cm}d^{ij}(x_{m_1}-x_{m_2})
\s\da_{x_{m_1}^i}\da_{x_{m_2}^j}\s.
\label{trin}
\qqq
Hence in the translationally invariant sector
the constant $D_0\sim m^{-\xi}$ divergent as the IR regulator
$m\to0$ decouples from operators $\CM_n$. For $\kappa=0$
and $m=0$, \s$\CM_n$ turn into scaling operators $\CM_n^{\rm sc}$
of homogeneity degree $\xi-2$.
\vskip 0.3cm

The 2-point function equation
\qq
\CF_2(x_1,x_2)\s=\s\CM_2^{-1}\CC({_{x_1-x_2}\over^L})
\label{2pt0}
\qqq
may be rewritten (with the scaling form
$d_0^{ij}$ of $d^{ij}$ and
$r\equiv|x_1-x_2|\s$) as
\qq
(\CM_2\CF_2)(r)\s=\s 2\m r^{1-d}\da_r\s(Dr^{d-1+\xi}+\kappa
r^{d-1})\s\da_r\s\CF_2(r)\s=\s\CC({_r\over^L})
\label{2pt2}
\qqq
leading to the explicit solution
\qq
\CF_2(r)\s=\s{_1\over^2}\int
\limits_{r}^\infty{{\int\limits_0^\rho
\CC({\sigma\over L})\m\sigma^{d-1}\m d\sigma}\over
{D\rho^{d-1+\xi}+\kappa\m\rho^{d-1}}}\s\m d\rho\s.
\label{2pt}
\qqq
The integration constants were chosen so that
$\da_r\CF_2(0)=0$ and $\CF_2(\infty)=0$ which is equivalent
to the use of the Green function kernel for $\CM_2^{\m-1}$
in eq.\s\s(\ref{2pt0}). Such a choice describes
the stationary 2-point function obtained from a localized
initial distribution of $T$ by waiting long enough.
It is trivial to analyze the integral in eq.\s\s(\ref{2pt})
explicitly and to see that, first,  the $\kappa\to 0$
limit of $\CF_2$ exists and, second, that
\qq
\CF_2(r)\ \s\smash{\mathop{=}\limits_{{\kappa=0\atop L\ {\rm large}}}}
\ \s A_{_\CC}\s L^{2-\xi}\s-\s{_{\CC(0)}\over^{2Dd(2-\xi)}}
\s r^{2-\xi}\ +\ \CO(L^{-2})\s.
\label{2pt1}
\qqq
In particular, for the 2$^{\s\rm nd}$ structure function
of $T$, we obtain
\qq
S_2(r)\s=\s2\CF(0)-2\CF_2(r)\s=\s{_{\CC(0)}\over^{Dd(2-\xi)}}
\s r^{2-\xi}\ +\ \CO(L^{-2})
\non
\qqq
in agreement with the naive dimensional analysis
of eq.\s\s(\ref{2pt0}). By similar dimensional analysis
of eq.\s\s(\ref{2np}), we may expect that $S_{2n}(r)\sim
r^{n(2-\xi)}$ for large $L$. This is the Kolmogorov-type
prediction and it is incorrect for $n>1$. The first hint
of what may go wrong with the dimensional argument may be
already seen in eq.\s\s(\ref{2pt1}) where on the right
hand side, beside the scaling contribution
$-\m{{\CC(0)}\over{2Dd(2-\xi)}}\s r^{2-\xi}\s$, \s there
appears the constant $\s A_\CC\s L^{2-\xi}\s$
diverging when $L\to\infty$. Any divergent
contributions to $\CF_2$ have to be annihilated by $\CM_2$
since the right hand side of eq.\s\s(\ref{2pt2}) is
regular when $L\to\infty$ and clearly constants are zero
modes of $\CM_2$.
\vskip 0.3cm

Similarly, contributions annihilated by $\CM_{2n}$
may appear in the $(2n)$-point correlation function $\CF_{2n}$.
A more refined analysis shows that, for sufficiently
small $\xi$ and for fixed $x_1,\dots,x_{2n}$,
the $\kappa\to0$ and $m\to0$ limits
of $\CF_{2n}$ exist and are dominated
for large $L$ by the contributions of the scaling
zero modes of $\CM_{2n}^{\rm sc}$:
\qq
\CF_{2n}({\bf x})\ \s\smash{\mathop{=}\limits_{\nu,m=0\atop
L\ {\rm large}}}\ \s A_{_{\CC,2n}}\s L^{n(2-\xi)-\zeta_{2n}}
\s\s\CF^0_{2n}({\bf x})\s+\s\CO(L^{-2+\CO(\xi)})
\ +\ \s.\ .\ .\ .
\label{2np}
\qqq
where \s$\CF^0_{2n}\s$ is a zero mode of $\CM_{2n}^{\rm sc}$
of the homogeneity degree $\zeta_{2n}$ given by eq.\s\s(\ref{zeta})
and the dots denote terms which do not depend
on one of the variables $x_m$ and consequently do not
contribute to the structure function $S_{2n}$. \s The amplitudes
$A_{\CC,2n}$ are non-universal in the sense that they
depend on the shape of the forcing covariance $\CC$.
The zero mode
\qq
\CF^0_{2n}({\bf x})\s\ =\s\sum\limits_{\rm permutations}
\hs{-0.4cm}(x_1-x_2)^2\s\cdots\s(x_{2n-1}-x_{2n})^2\ \ +\s\
\CO(\xi)\ +\ \s.\ .\ .\ .
\non
\qqq
and reduces for $\xi=0$ to a polynomial zero mode of the
$(2nd)$-dimensional Laplacian.
\vs 0.3cm

The essential tool in arriving at the result (\ref{2np})
is the use of the Mellin transform of eq.\s\s(\ref{2npt})
with the scaling operator $(\CM_{2n}^{\s\rm sc})^{-1}$,
\s supplemented by the perturbative expansion of the
scaling zero modes of $\CM_{2n}^{\rm sc}$ in powers
of $\xi$. The analysis has a renormalization group flavor.
The perturbative argument is applied to the single scale
problem in which the differential operator
$\CM_{2n}^{\rm sc}$ restricted to scaling functions
of a given homogeneity degree is analyzed. Such an operator
has a discrete
spectrum. The perturbative zero mode information is then
plugged into the inverse Mellin transform which assembles
different homogeneity degrees. More exactly, the scaling
zero modes enter the residues of poles of the
Mellin transform of the Green function
$(\CM_{2n}^{\rm sc})^{-1}$. The above analysis
should be contrasted with the direct perturbative expansion
of $\CF_{2n}$ in powers of $\xi$ which requires perturbative
treatment of the Green function of $\CM_{2n}^{\rm sc}$,
an operator with a continuous spectrum in $L^2$.
The latter expansion is plagued by logarithmic
divergences (proportional to powers of $\log{L}$)
which are resummed on the right hand side of (\ref{2np}).
Still, the anomalous $\CO(\xi)$ contribution to $\zeta_{2n}$
may be extracted from the coefficient of $\log{L}$
in the $\CO(\xi)$ term of $\CF_{2n}$
proportional to the integral
\qq
\int\delta_{p,k}\delta_{q,k}\s\s\s{\ee^{\m i\m(p\cdot x+
k\cdot y+q\cdot z)/L}\over p^2+k^2+q^2-k\cdot(p+q)}
\s\s\s{p\cdot q-{(p\cdot k)\m(q\cdot k)\over k^2}\over
|k|^d\s p^2\s q^2}\s\s\s\hat\CC(p)\s\s\hat\CC(q)\s\s\s\dd p\s
\dd k\s\dd q
\non
\qqq
where \s$\delta_{p,k}f(p)\equiv f(p+k)-f(p)\m$. \s
We shall unravel the renormalization group underlying
the exponentiation of the logarithmic divergences
in the perturbative treatment in powers of $\xi$
of the passive scalar in the next section.
\vskip 0.2cm

The correlation functions $\CF_{2n}$ are not smooth
at coinciding points even for $\kappa>0$. Nevertheless
\qq
\lim\limits_{y\to x}
\ (\nabla T)(x)\cdot(\nabla T)(y)\s\equiv\s\epsilon(x)
\non
\qqq
exists inside the correlations for $\kappa>0$ and defines
the dissipation field. In particular, the mean dissipation
rate \s$\bar\epsilon\equiv\langle\epsilon(x)\rangle
={_1\over^2}\m\CC(0)\s$, \s as may be easily seen from
eq.\s\s(\ref{2pt}). $\epsilon(x)$ does not vanish when
$\kappa\to0$ but is given by the dissipative anomaly
\qq
\lim\limits_{\kappa\to0}\ \epsilon(x)\ =\
\lim\limits_{y\to x}\ d^{ij}(x-y)\s(\da_{x^i}T)(x)
\s\m(\da_{y^j}T)(y)
\label{disan}
\qqq
holding inside correlation functions. The result (\ref{2np})
permits to infer the scaling behavior (\ref{strf},\ref{zeta})
of the structure functions and, together with the
dissipative anomaly (\ref{disan}), also the inertial
range scaling of the correlations involving the dissipation
field. One obtains, for example,
\qq
\langle\s\epsilon(x)\s\m\epsilon(y)\s\rangle-\bar\epsilon^2
\ \sim\ r^{\zeta_4-2(2-\xi)}\s.
\non
\qqq
\vskip 0.9cm

\no {\large\bf{LECTURE 4}}
\addtocounter{section}{1}
\vskip 0.5cm

It has been realized a long time ago \cite{Nel74} that there
exists a similarity between the behavior of the
2-point correlation functions of a nearly critical
statistical-mechanical systems and of the Fourier
transform of the equal-time velocity correlators
in a turbulent flow. Both have a scaling regime of
power-law decay followed by much stronger decay
at infinity. One may then establish the following
dictionary \cite{RoS78}\footnote{see also \cite{EyinkG}
for a more recent discussion and more references}
\qq
\cr
\hbox to 6 cm{{\bf critical phenomena}\hfill}\quad\quad\quad
\hbox to 5 cm{{\bf turbulence\hfill}}\cr\cr
\hbox to 6 cm{UV cutoff \hfill}\quad\quad\quad
\hbox to 5 cm{integral scale\hfill}\cr
\hbox to 6 cm{inverse correlation length\hfill}\quad\quad\quad
\hbox to 5 cm{viscous scale\hfill}\cr
\hbox to 6 cm{$T-T_c$\hfill}\quad\quad\quad
\hbox to 5 cm{viscosity $\nu$\hfill}\cr
\hbox to 6 cm{scaling regime\hfill}\quad\quad\quad
\hbox to 5 cm{inertial range\hfill}\cr
\hbox to 6 cm{anomalous conservation laws\hfill}\quad\quad\quad
\hbox to 5 cm{dissipative anomaly\hfill}\cr
\non
\qqq
This suggests that, very roughly, the turbulant phenomena
look like critical phenomena, provided that we invert
the scales interchanging short and long distances
or the position and the Fourier spaces. Were this true,
the short-distance universality of the critical phenomena
(independence of the long-distance behavior of the
microscopic details of the system) should be accompanied
by the long-distance universality in turbulence (insensitivity
of the short-distance behavior to boundary
effects or/and details of the energy injection).
\vskip 0.3cm

The right tool to study the scaling properties of the critical
phenomena and to establish their short-distance universality
has been provided by the Kadanoff-Wilson renormalization
group (RG) \cite{Kadan,WilsK}. Loosely speaking,
the RG idea is to look at the system
from further and further away so that its microscopic
details are eventually wiped out and many microscopically
different models start looking the same. By analogy,
it seems \cite{Nel75,RoS78,MDeD78} that the turbulent
systems require an inverse renormalization group (IRG)
analysis. By examining them through a stronger
and stronger magnifying glass, we would loose their
large-scale details from the vision range and should discover
a short-distance similarity of different turbulent
cascades. The presence of a finite but large correlation
length in the nearly critical systems beyond which
there is a crossover to the high temperature regime
would then correspond to the presence of the short viscous
scale in high Reynolds number flows beyond which
the dissipative regime sets in.
\vskip 0.3cm

RG had an enormous success in explaining critical
phenomena \cite{WNobel,Pfeuty}. Why is it then
that IRG never developed beyond the level of a vague
idea? Is the analogy between the critical
phenomena and turbulence too naive and missing totally
the essential points? In the author's opinion the reason
is different. RG is not a universal key
to all problems as it is sometimes thought. Its effective use
requires a correct choice of RG transformations and that, in turn,
requires a good understanding of physics of the system.
Similarly, an IRG-type analysis of turbulence will require
a deep use of knowledge of physics of turbulence.
To provide an argument for such a thesis, we shall show
that the IRG idea allows to systematize the analysis
of Kraichnan's passive scalar described in the previous
lecture and opens a possibility to extend it to more
complicated systems.
\vskip 0.4cm

Let us start by a short reminder of how one may perform
a RG analysis of the long-time long-distance asymptotics of
the (nearly) critical dynamics described by the stochastic
evolution equation
\qq
\da_t T\s=\s D\Delta T\s-\s{_{\delta\CV(T)}\over^{\delta T}}\s
+\s f
\label{ee11}
\qqq
with $\CV(T)=\int V(T(t))\m dt$ and the Gaussian noise
$f$ with mean zero and covariance
\qq
\langle\s f(t,x)\s\m f(s,y)\s\rangle\s=\s
\delta(t-s)\s\m\CC({x-y})\s.
\non
\qqq
The corresponding MSR action is
\qq
S(R,T)\s=\s S_0(R,T)\s+\s S_1(R,T)
\non
\qqq
where
\qq
&&S_0(R,T)=
i\hs{-0.1cm}\int\hs{-0.1cm}R(t,x)\m(\da_t T-\nu\Delta T)(t,x)
\s dt\s dx\m+\m{_1\over^2}\hs{-0.1cm}\int\hs{-0.1cm}
R(t,x)\m\CC(x-y)\m R(t,y)\s dt\s dx\s dy\m,\cr
&&S_1(R,T)= i\hs{-0.1cm}\int\hs{-0.1cm}R(t,x)\s{_{\delta\CV(T)}
\over^{\delta T(t,x)}}\s dt\s dx\s+\s{_1\over^2}\int
{_{\delta^2V(T(t))}\over^{\delta T(t,x)\delta T(t,x)}}\s dt\s dx\s,
\non
\qqq
see eq.\s\s(\ref{MSR1}). The $S_0$ part of the action may be
used to define the Gaussian "measure"
\qq
d\mu_{_G}(R,T)\ =\ \ee^{\m- S_0(R,T)}\s\s
DR\s\m DT\s\bigg/\s{\rm norm.}
\non
\qqq
with covariance $G$ given by the 2-point functions of the forced
diffusion (\ref{co},\ref{co1}).
\vskip 0.3cm

Let $R,\s T$ and
$\tilde R,\s\tilde T$ be two copies of Gaussian fields
distributed with measure $d\mu_G$. We shall decompose
\qq
&&R\s=\s\tilde R_{1/\la}\s+\s\rho\s,\cr
&&T\s=\s\tilde T_{1/\la}\s+\s\tau\s
\label{deco}
\qqq
demanding that $\rho,\s\tau$ be Gaussian fields independent
of $\tilde R,\s\tilde T$. The rescalings of $\tilde R,\s\tilde T$
are as in eq.\s\s(\ref{resc1}), i.e.\s\s they serve to exhibit
the long-time long-distance scaling of the linear forced
diffusion. By assumption, we have the factorization
\qq
d\mu_{_G}(R,T)\s=\s d\mu_{_G}(\tilde R,\tilde T)\s\s
d\mu_{_{\Gamma_\la}}(\rho,\tau)\s.
\label{fact}
\qqq
The 2-point functions of
$\rho,\s\tau$ building the covariance $\Gamma_\la$
are the differences of the 2-point functions of
$R\s,T$ and $\tilde R_{1/\la},\s\tilde T_{1/\la}$, very
much in the spirit of the Pauli-Villars regularization. Explicitly,
\qq
\langle\s\tau(t_1,x_1)\s\m\tau(t_2,x_2)\s\rangle\s=\s
\int\ee^{\m\vert t_1-t_2\vert\m D\m k^2\s-\s i\m k\cdot(x_1-x_2)}
\s\s{_{\hat\CC(k)-\hat\CC(\la k)}\over^{2Dk^2}}\s\s\dd k\s.
\label{cop1}
\qqq
i.e. it is the high-momentum part of the covariance $\langle T\s
T\rangle$. The 2-point functions involving $\rho$ vanish
(recall that $\langle R\m R\rangle=0$ and $\langle T\m R\rangle$
is scale-invariant) but we shall keep $\rho$ in the formulae
which will be later applied in situations with $\rho\not=0$.
The decomposition (\ref{deco}) of $T$ is into the
low-momentum part $\tilde T_{1/\la}$ and the high-momentum
fluctuation $\tau$ and allows to define the effective
interactions
\qq
\ee^{\m-S_\la(\tilde R,\tilde T)}\ =\ \int\ee^{\m
-S_1(\tilde R_{1/\la}+\rho,\s\tilde T_{1/\la}+\tau)}
\ d\mu_{_{\Gamma_\la}}(\rho,\tau)
\label{efint}
\qqq
by integrating out the high-momentum fluctuations from
the Boltzmann factor $\ee^{-S_1}$. The RG transformations
\s$\CR_\la:\m S_1\mapsto S_\la$ have a semigroup property,
$\CR_\la\circ\CR_{\la'}=\CR_{\la\la'}$. We may
also integrate out the high-momentum fluctuations
in the insertions $F_1(R,T)$ into the MSR functional
integral, defining the effective insertions by
\qq
F_\la(\tilde R,\tilde T)\s=\s
\int F_1(\tilde R_{1/\la}+\rho,\m\tilde T_{1/\la}+\tau)\
\ee^{\m-S_1(\tilde R_{1/\la}+\rho,\s\tilde T_{1/\la}+\tau)}
\ d\mu_{_{\Gamma_\la}}(\rho,\tau)\s\bigg/\s
\ee^{-S_\la(\tilde R,\tilde T)}\m.\hs{0.7cm}
\label{efins}
\qqq
Note that the transformation $F_1\mapsto F_\la$
is really
the derivative $d\CR_\la(S_1)$ of the semigroup $\CR_\la$
and that the RG transformations preserve
the expectation values:
\qq
\langle\s F_1\s\rangle_{_{S_1}}\ =\ \langle\s F_\la\s
\rangle_{_{S_\la}}
\label{pres}
\qqq
where \s$\langle\s F\s\rangle_{_{S_\la}}\s\equiv\s
\int F\s\m\ee^{-S_\la}\s/\s\int\ee^{-S_\la}\s.\s$
\vskip 0.3cm

The vanishing interaction $S_1=0$ corresponds
to a (Gaussian, trivial) fixed point of the semigroup
$\CR_\la$. For non-trivial $S_1$, the simplest situation
occurs if under the action of $\CR_\la$ it converges
to a (possibly trivial) fixed point:
\qq
S_\la\ \ \smash{\mathop{\longrightarrow}
\limits_{\la\to\infty}}\ \ S_*\s.
\non
\qqq
The study of the long-distance asymptotics
of the correlation functions reduces then to the search
of the corresponding scaling fields. Let us explain
the latter concept. Let \s$F(R,T;{\bf x})$ be a
functional of $R,\s T$ explicitly dependent on
a sequence of space-points
${\bf x}$. For example, we may take \s$F(R,T;{\bf x})=
\prod\limits_i T(t,x_i)\s$. Suppose further, that for some
exponent $\zeta_*$
\qq
\la^{-\zeta_*}\s\m (F(\la{\bf x}))_\la
\ \ \smash{\mathop{\longrightarrow}
\limits_{\la\to\infty}}\ \ F_*({\bf x})\s.
\label{const}
\qqq
Then, tautologically,
\qq
(F_*({\bf x}))_\la\s=\s\la^{\zeta_*}\s\m F_*(\la^{-1}{\bf x})
\non
\qqq
if in the computation of the effective insertion
on the left hand side we use the fixed point interaction
$S_*$. Hence the name: scaling field for $F_*({\bf x})$.
Besides, in view of eq.\s\s(\ref{pres}),
\qq
\lim\limits_{\la\to\infty}\ \la^{-\zeta_*}\s\m\langle\s
F(\la{\bf x})\s\rangle_{_{S_1}}\ =\ \s\langle\s F_*({\bf x})\s
\rangle_{_{S_*}}
\non
\qqq
giving the long-distance asymptotics of
\s$\langle F({\bf x})\rangle_{_{S_1}}\s$ (if the fixed-point
expectation on the right hand side does not vanish).
\vskip 0.3cm

The first information about the RG flow $\CR_\la$ may be
obtained by studying its linearization around
the Gaussian fixed point. In fact for functionals
$S_1$ polynomially dependent on $R,\m T$ with local scaling
kernels
\qq
S_\la\s=\s d\CR_\la(0)\s S_1\s+\s\CO(S_1^{\m2})\s=\s
\la^{[S_1]}\s\m S_1\s+\s{\rm lower\ order\ polyn.}
\s+\s\CO(S_1^{\m2})
\label{lRG}
\qqq
where the (long distance) dimension $[S_1]$ of $S_1$
is calculated
additively with the use of the following table
\qq
[x]=1\s,\quad[t]=2\s,\quad[T]=1-{_d\over^2}\s,
\quad[R]=-1-{_d\over^2}\s.
\non
\qqq
Small irrelevant interactions with $[S_1]<0$
should then die out under the iterated RG
transformations\footnote{this may require
fine tuning of the lower order terms in $S_1$}
resulting in the convergence of $S_\la$ to
the trivial fixed point and in the long-time
long-distance asymptotics of the correlations as in the
linear forced diffusion case. The fate
of relevant ($[S_1]>0$) or marginal ($[S_1]=0$)
interactions cannot be determined by the linear analysis
around the trivial fixed point and requires higher
order calculations which may show convergence
to a non-trivial fixed point situated
in a perturbative neighborhood of the trivial one.
For example, the nonlinearity with $\CV(T)\sim\int T^4$ in
eq.\s\s(\ref{ee11}), describing the Langevin-type dynamics
for the $\phi^4$ field theory, leads
to the interaction \s$\sim\int RT^3$ of dimension
$-1-{d\over2}+3-{{3d}\over2}+2+d=4-d$,
\s in agreement with the well known static
power counting rendering the nonlinearity
irrelevant above 4 dimensions. For $d<4$,
the RG flow is instead governed by a non-trivial
fixed point which seems accessible by a perturbative
$\epsilon$-expansion in powers of $\epsilon\equiv4-d$
\cite{WilsK}. An important aspect, essential for the validity
of the RG analysis in the above systems, is the approximate
locality of the effective interactions $S_\la$ in the position
space which means physically that no low-energy interacting
modes were removed from the system by integrating out
the short-distance fluctuations. Such locality (usually checked
only perturbatively) allows to separate the dominant exactly
local scaling contributions to $S_\la$ driving the RG dynamics
from the remainder strongly damped under the RG flow.
Technically, the separation is done by Taylor-expanding
the kernels in $S_\la$ in the Fourier space (the
approximate locality of the kernels in the position space
makes their Fourier transforms smooth).
\vskip 0.5cm

Suppose that instead of being interested in the long-time,
long-distance behavior of the nonlinear forced diffusion,
we want to study its short-time, short-distance asymptotics.
In particular, we would like to know how the nonlinearity
effects the UV asymptotics (\ref{lr2}) of the solutions
of the linear equation. In order to study this problem, we
may set up an (inverse) RG formalism in full analogy
with the one described above, with the only difference
that we shall use the field rescalings (\ref{resc2})
suitable for tracing the short-time, short-distance
asymptotics of the linear forced diffusion
instead of (\ref{resc1}) appropriate for the long-time
long-distance behavior. Repeating the decomposition
(\ref{deco}) leading to the factorization
(\ref{fact}) for the new rescaling (marked in the
notation by superscripts rather then subscripts),
we shall obtain the fluctuation $\tau$ covariance
\qq
\langle\s\tau(t_1,x_1)\s\m\tau(t_2,x_2)\s\rangle\s=\s
\int\ee^{\m\vert t_1-t_2\vert\m D\m
k^2\s-\s i\m k\cdot(x_1-x_2)}
\s\s{_{\hat\CC(k)\m-\la^{-d}\m\hat\CC(k/\la)}
\over^{2Dk^2}}\s\s\dd k\s.
\label{cop2}
\qqq
Note that $\hat\CC(k)-\la^{-d}\hat\CC(k/\la)$ is the
Fourier transform of the $\CC(x)-\CC(\la x)$ so
that it corresponds to the long distance part
of $\CC$. In other words, the decomposition
$T=\tilde T^{1/\la}+\tau$ is now into the short-distance
part and the long-distance fluctuation.
Repeating the definitions (\ref{efint}) and (\ref{efins})
for the new scaling, we obtain the IRG semigroup
$\CR^\la:\s S_1\equiv S^1\mapsto S^\la$ with
the derivative \s$d\CR^\la(S^1):\s F_1\equiv F^1\mapsto
F^\la$. If the effective long-distance interactions
$S^\la$ converge to a fixed point $S^*$, the study
of the short-distance asymptotics of the correlation
functions for the perturbed version of the forced
diffusion reduces to the search for the scaling
fields
\qq
\lim\limits_{\la\to\infty}\ \la^{\zeta^*}\s\m
(F({\bf x}/\la))^\la\ =\ \s F^*({\bf x})
\non
\qqq
for which
\qq
\lim\limits_{\la\to\infty}\ \la^{\zeta^*}\s\m
\langle\s F({\bf x}/\la)\s\rangle_{_{S^1}}\ =\ \s
\langle\s F^*({\bf x})\s\rangle_{_{S^*}}\s.
\non
\qqq
For small interactions $S^1$,
\qq
S^\la\s=\s d\CR^\la(0)\s S^1\s+\s\CO((S^1)^2)\s=\s
\la^{-[S^1]}\s+\s{\rm lower\ order\ polyn.}\s+\s\CO((S^1)^2)
\label{lIRG}
\qqq
where the (short distance) dimension $[S^1]$ is calculated
with the use of the new table
\qq
[x]=1\s,\quad[t]=2\s,\quad[T]=1\s,
\quad[R]=-1-d\s.
\non
\qqq
Note the change of the sign in the exponent of $\la$
in eq.\s\s(\ref{lIRG}) as compared to eq.\s\s(\ref{lRG})).
For example, nonlinearity $\CV(T)\sim\int(T)^4$
in eq.\s\s(\ref{ee11}) leading to the interaction
\s$\sim\int R\m T^3$ of dimension \s$-1-d+3+2+d=4\m$, \s i.e.
irrelevant by power counting. $\int R\m T^3$ couples, however,
to the unstable constant mode of $T$. Considering
instead the gradient-type nonlinearity with
$\CV(T)\sim\int(\nabla T)^4$ in eq.\s\s(\ref{ee11}),
avoiding coupling to the constant mode, one obtains
the interaction \s$\sim\int R\m\nabla(\nabla T)^3$
of dimension \s$-1-d-1+2+d=0\m$, \s i.e. marginal in all dimensions.
Hence the linearized IRG does not provide any simple hints
about the short-time, short-distance asymptotics of the nonlinear
forced diffusion. Besides, one should check
that the IRG effective interactions $S^\la$
possess in this case a Fourier space locality properties
which would allow to separate a finite number of scaling
contributions driving the IRG flow. It should be
also noticed that the covariance (\ref{cop2}), unlike
its RG counterpart (\ref{cop1}), is not positive
which may lead to non-perturbative complications in stabilizing
the IRG flow. We shall have nothing more to say about the
UV regime of the forced linear diffusion except repeating
that its control is an interesting open problem with
physical relevance. Below, we shall apply the IRG idea
to the passive scalar model of Lecture 3 with milder
nonlinearities and milder stability problems.
\vskip 0.3cm

The mutual relations of RG and IRG lead often to a confusion
stemming from the fact that the RG is also used to study
the short-distance asymptotics in field theories governed
by UV fixed points. Thus there are two contexts in which we
apply the standard RG: either we fix the UV cutoff and
study the long-distance behavior of the theory by observing
stabilization of the system under RG which lowers the
momentum cutoff (the statistical-mechanical context)
or we start with theories with a larger and larger momentum
cutoff and apply RG to lower it to a fixed value,
adjusting the parameters of the cutoff theories as to
obtain stabilization of the effective theories on
the fixed scale (the continuum limit or field theory context).
As is well known, the two contexts differ essentially only
by a straightforward rescaling and RG used in both of them
integrates out the degrees of freedom corresponding to the
shortest distances present. Similarly, IRG may be
applied to systems forced on long distances in two contexts
differing essentially by a simple rescaling. Either we fix
the size of the system and the forcing scale (the IR cutoffs)
and study the short-distance behavior by trying to exhibit
stabilization under IRG which lowers the long-distance
cutoffs (the more common situation) or we consider larger
and larger systems forced at longer and longer distances
and apply IRG to lower the distance cutoff to a fixed
value hoping to see stabilization of the effective theories
on the fixed distance scale (the infinite-volume limit context).
Again, in both contexts the IRG transformations lower
the distance cuttoff. In a given situation whether RG or
IRG should be used depends on which leads to a finite
number of expanding directions in the interaction
space\footnote{it is {\it a priori} not excluded that
there are exotic systems which yield to both RG and IRG
analysis}. It seems that the limitation of the energy injection
to long distances condemns the RG analyses of turbulence
employing a modified forcing with a power spectrum and
the $\epsilon$-expansion techniques \cite{FNS,DeDM0,YaOrsz}.
\vskip 0.4cm

We would like to study via IRG the short-distance
asymptotics of the passive scalar (\ref{ps})
with the random velocity distributed
according to eqs.\s\s(\ref{covv}-\ref{covv1}).
Under the rescaling
\qq
v^\la(t,x)\s=\s\la^{\xi-1}\s v(t/\la^{2-\xi},\m x/\la)\s,
\label{vvre}
\qqq
\qq
\langle\s{v^\la}^i(t,x)\s\m{v^\la}^j(s,y)\s\rangle\s
=\s\delta(t-s)\s\bigg(2\la^\xi D_0\delta^{ij}\s-\s2\m
d_0^{ij}(x-y)\ +\ \CO(({_m\over^\la})^{2-\xi})\bigg)
\non
\qqq
for large $\la$. Hence $\langle v\m v\rangle$ exhibits
a scaling behavior at short distances, modulo a divergent
constant zero mode.
One of the problems with the passive scalar is that
the term $-\m v\cdot\nabla T$ couples $T$ to the unstable
constant mode of $v$. A possible solution is to replace
eq.\s\s(\ref{ps}) by its quasi-Lagrangian version
\cite{BeLv}
\qq
\da_t T\s=\s -\m (v-v_0)\cdot\nabla T\s+\s\kappa
\Delta T\s+\s f\s.
\label{psqL}
\qqq
where $v_0(t,x)\equiv v(t,0)$. This changes the stochastic
dynamics. The new dynamics may be interpreted as
describing the system in the frame moving with one
of the material points of the flow \cite{BeLv}.
Although eq.\s\s(\ref{psqL}) corresponds to different
dynamical correlations of $T$, the equal-time correlations
do not change \cite{GK0}. As in Lecture 3, one may
obtain for the latter the same equations (\ref{2npt})
with $\CM_n$ directly in the translation-invariant form
(\ref{trin}). Of course, instead of employing the new equation
(\ref{psqL}), we may keep the original one (\ref{ps}) changing
only the velocity covariance to
\qq
\langle\s v^i(t,x)\s\m v^j(s,y)\s\rangle\s=\s
-\m2\m\delta(t-s)\left(d^{ij}(x-y)-d^{ij}(x)
-d^{ij}(y)\right)\s.
\label{coqL}
\qqq
It will be more convenient to choose another $\langle v\m v\rangle$
covariance, which we shall call mixed,
\qq
\langle\s v^i(t,x)\s\m v^j(s,y)\s\rangle\s=\s
-\m\delta(t-s)\left(2\m d^{ij}(x-y)-d^{ij}(x)
-d^{ij}(y)\right)
\label{comi}
\qqq
which leads to the equations (\ref{2npt}) with $\CM_n$
replaced by
\qq
\CM'_n\ =\ -\sum\limits_{m=1}^n(\kappa\Delta_{x_m}
\m+\m d^{ij}(x_m)\s\da_{x_m^i}\da_{x_m^j})\ \s+\hs{-0.3cm}
\sum\limits_{1\leq m_1<m_2\leq n}\hs{-0.3cm}(2\m d^{ij}(x_{m_1}
-x_{m_2})-d^{ij}(x_{m_1})\ \cr
-d^{ij}(x_{m_2}))\s\da_{x_{m_1}^i}\da_{x_{m_2}^j}\s.\hs{0.6cm}
\non
\qqq
$\CM'_n$ again coincide with $\CM_n$ in the translational-invariant
sector so that the mixed distribution (\ref{comi}) of $v$
leads to the same equal-time correlators of $T$. The covariance
(\ref{comi}) is $\CO(\xi)$ for $\kappa=0$ and its use will
simplify the perturbative expansion in powers of $\xi$. Although
it lacks positivity, unlike the original covariance or the
quasi-Lagrangian one, this will not cause stability problems.
\vskip 0.4cm

We shall introduce another modification of eq.\s\s(\ref{ps})
by replacing it by
\qq
\da_t T\s=\s -\m v\cdot\nabla T\s+\s\kappa\Delta T\s
-\s\eta\m m^2T\s+\s f
\label{ps2}
\qqq
where $m$ is the same IR regulator that in eq.\s\s(\ref{covv1})
and $\eta>0$ is a small fixed constant. The addition
of the "mass term" $\sim m^2T$, which changes the long-distance
behavior is innocuous for the short-distance one, see below.
The expectation values for the passive scalar (\ref{ps2})
with the velocity distribution (\ref{comi}) may be generated
in the MSR formalism in the following way.
Introduce operator
\qq
M_m\s=\s-\m d^{ij}_m(x)\s\da_{x^i}\da_{x^j}\s+\s\eta m^2\s.
\non
\qqq
$M_m$ is a generator of inhomogeneous
superdiffusion (on distance scales $\ll m^{-1}$) and
$\ee^{-tM_m}$ describes the dynamics of the 1-point
function of the passive scalar with velocity covariance
(\ref{comi}).
Let
\qq
&&S^0(R,T)\s=\s i\int R(t,x)\m(\da_t-M_m)T(t,x)\s dt\m dx\s
+\s{_1\over^2}\int R(t,x)\s\m\CC(x-y)\s R(t,y)
\s dt\m dx\m dy\m,\cr
&&S^1(R,T,v)\s\s=\s i\int R(t,x)\s v(t,x)\cdot\nabla T(t,x)\s\s
dt\s dx\s,\cr
&&{S'}^1(R,T,v)\s\s=\s-\m i\m\kappa\int R(t,x)\m\Delta T(t,x)\s\s
dt\s dx\s,\cr
&&{S''}^1(R,T)\s=\s i\int R(t,x)\m M_mT(t,x)\s\m dt\s dx\s.
\non
\qqq
Let $d\mu_{_{D'_m}}(v)$ denote the Gaussian measure with
the mixed covariance (\ref{comi}).
The equal time $(2n)$-point function $\CF_{2n}({\bf x})$ of
the scalar $T$ may be represented as the MSR functional
integral
\qq
\CF_{2n}({\bf x})\ =\ \int\prod\limits_{i=1}^{2n} T(t,x_i)
\ \ee^{-S(R,T,v)}\s\s DR\s\m DT\s\s d\mu_{_{D'_m}}(v)
\ \bigg/\ {\rm norm}.
\label{fir}
\qqq
where $S\equiv S^0+S^1+{S'}^1+{S''}^1$. We have included the
$-i\int R\m M_m T$ term into free action $S^0$ compensating
it by the ${S''}^1$ term treated as an interaction. The role
of ${S''}^1$ is to remove the "tadpole" contractions
$\langle(v\cdot\nabla)^2\rangle$
in $\langle(S^1)^2\rangle$. The latter are instead resummed
into the free $\langle T\m R\rangle=i(\da_t-M_m)^{-1}$
propagators originating from $S^0$. Reading the field
short-distance dimensions from $S^0$ and, for $v$,
from the short-distance
scaling of the $\langle v\m v\rangle$ covariance, we obtain
\qq
[x]=1\s,\quad[t]=2-\xi\s,\quad[T]=1-{_1\over^2}\xi\s,
\quad[R]=-1-d+{_1\over^2}\xi\s,\quad[v]=\xi-1\s.\hs{0.3cm}
\label{tble}
\qqq
Note that the Kolmogorov value of the velocity exponent
$\xi={4\over3}$ is obtained
by equating the dimensions of $T$ and $v$ which is encouraging
in view of the fact that the $v\cdot\nabla T$ term of the
passive scalar equation is replaced by $v\cdot\nabla v$
in the NS equation. The table (\ref{tble}) gives
\qq
[S^1]=0\s,\quad[{S''}^1]=0\s,\quad[{S'}^1]=-\xi\s.
\non
\qqq
\vskip 0.3cm

We may attempt an IRG analysis of the short-distance
behavior of the correlations $\CF_{2n}$, basing the
IRG transformations on the Gaussian measure
\s$d\mu_{_G}(R,T,v)\sim\s\ee^{-S^0(R,T)}\s DR\s DT
\s\m d\mu_{_{D'}}(v)\m$ and rescalings corresponding
to the table of dimensions (\ref{tble}),
\qq
&&R^\la(t,x)\s=\s\la^{-1-d+\xi/2}\s R(t/\la^{2-\xi},\m x/\la)\s,\cr
&&T^\la(t,x)\s=\s\la^{1-\xi/2}\s T(t/\la^{2-\xi},\m x/\la)\s
\label{nrsc}
\qqq
and $v^\la(t,x)$ as in eq.\s\s(\ref{vvre}).
We could hope that
the effective interactions $S^\la$ obtained from the marginal
terms $S^1+{S''}^1$ tend to a fixed point, with the relevant
${S'}$ term with very small $\kappa$ destabilizing
the IRG trajectory only at very short distances causing
eventually a crossover to the dissipative regime.
A closer analysis of the effective interactions
based on the $\xi$ expansion shows however a lack
of convergence of $S^\la$ defined this way
to a fixed point \cite{BGHK}. It appears that
an infinite number of relevant terms is generated
which destabilize the effective trajectory.
In such situation, it may seem that IRG
fails to predict the short-distance scaling
of the scalar correlations in the inertial range.
How is it possible then that we still were able to
control this scaling for small $\xi$, as discussed in
Lecture 3?
\vskip 0.2cm

There appears to exist a simple solution to the above paradox.
The idea is to exclude the ${_1\over^2}\int R\CC R$ term
from the free action $S^0$ expanding in the functional
integral (\ref{fir}) \s$\ee^{-{1\over2}\int R\CC R}
=\sum\limits_{n=0}^\infty A_n\s\s F_{_\CC}^{\m n}\s$
with \s$A_n\equiv{_{(-1)^n}\over^{2^n\m n!}}\s$ and
\s$F_{_\CC}(R)=\int R(t,x)\s\CC(x-y)\m
R(t,y)\s dt\s dx\s dy\s$. \s Note that only the $n^{\m\rm th}$
term of the expansion gives a non-zero contribution to
(\ref{fir}) (the number of the $T$ and $R$ insertions
must be equal now). The leftover free action
\s${S'}^0=i\int R\m(\da_t-M_m)T\s$ leads to the Gaussian
measure with the 2-point functions $\langle R\m R\rangle=0=
\langle T\m T\rangle$ and \m$\langle T(t_1)\m R(t_2)\rangle=
i\m(\da_t-Mm)^{-1}(t_1,t_2)=
i\m\theta(t_1-t_2)\s\ee^{-(t_1-t_2)M_m}\m$. It will be
more convenient to introduce a cutoff version
$d\mu_{_{\Gamma^\La_m}}(R,T)$ of this measure
with the 2-point functions
\qq
&&\langle\s R\s R\s\rangle\s=\s0\s,\quad\quad
\langle\s T\s T\s\rangle\s=\s0\s,\cr\cr
&&\langle\s T(t_1,x_1)\s\m R(t_2,x_2)\s\rangle\s=\s
i\m(\da_t+M_m)^{-1}(t_1,x_1;\m t_2,x_2)\s
-\s i\m(\da_t+M_{\La m})^{-1}(t_1,x_1;\m t_2,x_2)\cr\cr
&&\hs{3.74cm}=\s i\m\theta(t_1-t_2)\left(\ee^{-(t_1-t_2)M_m}-
\m\ee^{-(t_1-t_2)M_{\La m}}\right)(x_1,x_2)\s.
\non
\qqq
For $\La\to\infty$,
\m$\langle T(t_1)\m R(t_2)\rangle\m$ tends
to\footnote{the $\eta\m m^2$ term in $M_n$ plays here
the crucial role}
\s$i\m\theta(t_1-t_2)\s\ee^{-(t_1-t_2)M_m}\s$ except for
$t_1=t_2$ for which it always vanishes. Finite $\La$
introduces a short-time, short-distance cutoff into the heat kernel
\s$\ee^{-(t_1-t_2)M_m}\s$. \s The functional-integral
expression (\ref{fir}) for $\CF_{2n}$ may be now rewritten
as
\qq
\CF_{2n}({\bf x})\ =\ A_n\ \lim\limits_{\La\to\infty}\ {
\int\prod\limits_{i=1}^{2n} T(t,x_i)\ F_{_\CC}(R)^{{n}}
\ \ee^{-(S^1+{S'}^1)(R,T,v)}\ d\mu_{_{\Gamma^\La_m}}(R,T)\
d\mu_{_{D'_m}}(v)\over{\rm norm}.}\cr\cr
\equiv\ A_n\ \langle\s\prod\limits_{i=1}^{2n}T(t,x_i)\s\s
F_{_\CC}(R)^{{n}}\s\rangle_{_{S^1+{S'}^1}}\s.\hs{0.5cm}
\label{2npp}
\qqq
The role of the short-distance cutoff $\La$ is the same
as that of the ${S''}^1$ term before. It excludes
the tadpole contractions $\langle(v(t)\cdot\nabla)^2\rangle$
in $\langle(S^1)^2\rangle$ now forbidden because
the cutoff $\langle T(t_1)\m R(t_2)\rangle$ propagator,
unlike its $\La=\infty$ version, vanishes at equal times
for all $\La$.
Note how eq.\s\s(\ref{2npp}) works for the 2-point function.
Expanding on the right hand side
$\ee^{-S^1+{S'}^1}$ into the power series, computing
the Gaussian expectations and passing to the $\La\to\infty$
limit, one obtains the Neuman series
for ${\CM'_2}^{\hs{-0.05cm}-1}\CC$ resulting from treating
as a perturbation the second line in the identity
\qq
\CM'_2\s&=&\s(M_m)_{x_1}\s+\s(M_m)_{x_2}\cr\cr
&+&\s(2d^{ij}(x_1-x_2)-d^{ij}(x_1)-d^{ij}(x_2))\s\da_{x_1^i}
\da_{x_2^j}\s-\s\kappa\Delta_{x_1}\s-\s\kappa\Delta_{x_2}\s.
\non
\qqq
The higher-point function formula works similarly.
\vskip 0.3cm

In order to generate the IRG transformations,
we shall follow the general rules described before.
Introducing the Gaussian fields $\tilde R,\m\tilde T,\m\tilde v$
distributed with the measure $d\mu_{_{\Gamma^{\La/\la}_m}}
(\tilde R,\tilde T)\s d\mu_{_{D'_m}}(\tilde v)$ and decomposing
\qq
R\s=\s\tilde R^{1/\la}\s+\s\rho\s,\quad\
T\s=\s\tilde T^{1/\la}\s+\s\tau\s,\quad\
v\s=\s\tilde v^{1/\la}\s+\s w\s
\non
\qqq
and, for the measures,
\qq
d\mu_{_{\Gamma^{\La}_m}}
(R,T)\s=\s
d\mu_{_{\Gamma^{\La/\la}_m}}
(\tilde R,\tilde T)\s
\ d\mu_{_{\Gamma^{\la}_m}}(\rho,\tau)\s,\quad\ \
d\mu_{_{D'_m}}(v)\s=\s d\mu_{_{D'_m}}(\tilde v)\
d\mu_{_{\delta^\la_m}}(w)\hs{0.3cm}
\non
\qqq
with
\qq
\delta^\la_m\s=\s D'_m-D'_{\la m}\s,
\non
\qqq
we shall define effective interactions $S^\la=\CR^\la(S^1)$
by integrating the long distance fluctuations
$\rho,\m\tau,\m w$
in the Gibbs factor $\ee^{-S^1}$,
\qq
\ee^{-S^\la(\tilde R,\tilde T,\tilde v)}\ =\
\int\ee^{-S^1(\tilde R^{1/\la}+\rho,\m\tilde T^{1/\la}+\tau,
\m\tilde v^{1/\la}+w)}\ d\mu_{_{\Gamma^{\la}_m}}(\rho,\tau)\s\s
d\mu_{_{\delta^\la_m}}(w)\s.
\non
\qqq
The effective insertions will be given by the linearization
$d\CR^\la$ of the above IRG flow for interactions, compare
eqs.\s\s(\ref{efint},\ref{efins}). \m Denote
\s$(T(t,x_1)-T(t,x_2))^{2n}\s\equiv\s
F_{2n}(T;{\bf x})\s$.
\vskip 0.4cm

Our main claims are as follows:
\vskip 0.2cm
\noindent 1.\ \ For the molecular diffusivity $\kappa$ set to zero,
the effective interactions $S^\la$ tend to the (non-trivial)
limit $S^*$ as $\la\to\infty$,
at least order by order in the expansion in powers of $\xi$.
At a fixed order, $S^*(\tilde R,\tilde T,\tilde v)$ is given
by an explicit integral expressions with kernels approximately
local (i.e.\s\s fast decaying) in the Fourier space.
\vskip 0.1cm
\noindent 2.\ \ The limit
\qq
\lim\limits_{\la\to\infty}\ \la^{n(2-\xi)}\s\s
(F_{2n}({\bf x}/\la))^\la\s\equiv\s F_{2n}^{\s*}({\bf x})
\non
\qqq
exists and is a scaling field of dimension
$n(2-\xi)$ of the fixed-point theory.
\vskip 0.1cm
\noindent 3.\ \ The limit
\qq
\lim\limits_{\la\to\infty}\ \la^{\zeta_{2n}-n(2-\xi)}
\s\s (F_{_\CC}^n)^\la\s\equiv\s(F_{_\CC}^n)^*
\non
\qqq
exists and is a scaling field of dimension $\zeta_{2n}-n(2-\xi)=
-{2n(n-1)\over d+2}\xi+\CO(\xi^2)$, see eq.\s\s(\ref{zeta}).
\vskip 0.1cm
\noindent 4.\ \  Finally, the limit
\qq
\lim\limits_{\la\to\infty}\ \la^{\zeta_{2n}}
\s\s(F_{2n}({\bf x}/\la)
\s F_{_\CC}^n)^\la\s\equiv\s F_{_{2n,\CC}}^{\s\m*}({\bf x})
\non
\qqq
exists and is a scaling field of dimension $\zeta_{2n}$.
\vskip 0.2cm

\noindent The last three results have been established to the
first two non-trivial orders in $\xi$. We shall not discuss
here the details of the perturbative analysis referring
an interested reader to \cite{BGHK}. Let us only mention
the role of the Mellin transform of the kernels entering
the effective interactions or insertions in separating
the contributions with the lowest dimensions from
the reminders strongly damped by the IRG flow.
\vskip 0.3cm

What are the implication of the above results?
First, note that, in view of eq.\s\s(\ref{2npp}), the relation
\qq
\lim\limits_{\la\to\infty}\ \la^\zeta_{2n}\s\s\langle
\s F_{2n}({\bf x}/\la)\s F_{_\CC}\s\rangle_{_{S^1}}\s=\s
\langle\s F_{_{2n,\CC}}^{\s\m*}({\bf x})\s\rangle_{_{S^*}}
\label{sf3}
\qqq
implies the anomalous scaling (\ref{strf}): unlike
the expectations $\langle F_{2n}^{\s*}\rangle_{_{S^*}}$ and
$\langle (F_{_\CC}^n)^*\rangle_{_{S^*}}$ which involve
unequal numbers of $R$ and $T$ fields, the right hand side
of eq.\s\s(\ref{sf3}) does not vanish. The even more
interesting observation is that the anomalous dimensions
$\zeta_{2n}-n(2-\xi)$ are carried by the scaling fields
$(F_{_\CC}^n)^*$ which are relevant (i.e. of negative dimension)
for small $\xi$ and which correspond to the composite operators
$F_{_\CC}^n$. The scaling zero modes of the operators
$\CM_{2n}$ which played the crucial role in
obtaining (\ref{strf}) enter the kernels
in the explicit formulae for $(F_{_\CC}^n)^*$.
Recall that in statistical mechanics
or field theory local composite operators are produced
by multiplying fields localized at the same space-point.
In the spirit of the scale inversion discussed at the beginning
of this lecture, composite fields in turbulence
should be obtained by multiplying fields localized
at the same wavenumber in the Fourier space. Since
the forcing covariance $\CC$ is almost a constant
in the inertial range, \s$F_{_\CC}\cong\CC(0)\int
\hat R(t,k)^2|_{_{k=0}}\s dt\s$ so that the
$F_{_\CC}^{\m n}$ insertions are almost local
in the wavenumber space. On the other hand, the scaling
fields $F_{2n}^{\m*}({\bf x})$ corresponding to the insertions
$(T(t,x_1)-T(t,x_2))^{2n}$ carry the Kolmogorov (normal) part
of the dimension of the structure functions. There are no
extra anomalous dimensions appearing in the scaling fields
$F_{_{2n,\CC}}^{\s\m*}({\bf x})$ which correspond to
the products of $F_{2n}({\bf x})$ and $F_{_\CC}^n$.
Their dimensions are just the sum of the dimensions of
$F_{2n}^{\m*}({\bf x})$ and of the composite scaling
fields $(F_{_\CC}^n)^*$. The IRG picture of the system
permits to analyze systematically such operator products
in the spirit of operator product expansions of the
long distance type (as contrasted with the field-theoretic
short-distance OPE's \cite{WilsZ} resulting from
the RG analysis).
Such long-distance OPE's lead to fusion rules
of the type studied in \cite{ProcLv}.
\vskip 0.2cm Up to now, we have ignored the
relevant ${S'}^1$ contribution to the MSR action
of the model, proportional to the molecular diffusivity
$\kappa$. As discussed above, it eventually causes
a crossover of the IRG trajectory from the vicinity
of the convective fixed point $S^*$ to another one
corresponding to the dissipative regime dominating
very short distances. It should be interesting to study
the crossover and the dissipative regime using IRG.
As for the role of the $\eta\m m^2 T$ term which we have
added in the passive scalar equation (\ref{ps2}),
note that its change gives rise to a term $\sim i\int R\m T$
in the MSR action of dimension $2-\xi$, i.e.\s\s irrelevant
at short distances. We could, in fact, analyze directly
the $\eta=0$ case, introducing $\eta>0$ only in the
intermediate IRG steps.

\vskip 0.3cm
\un{Summarizing}.\ \ The IRG analysis explains
the breakdown of the Kolmogorov scaling of the higher
structure functions of Kraichnan's passive
scalar for small $\xi$ as due to the appearance
of relevant composite scaling fields $(F_{_\CC}^n)^*$
with a multifractal spectrum of dimensions.
These fields do not destabilize the convergence
of the effective interactions $S^\la$ to the fixed point
$S^*$ since their appearance in $S^\la$ is forbidden
by the conservation law imposing the equality
of the numbers of
$R$ and $T$ fields in interactions (this was not
the case in our initial attempt to perform the
IRG analysis which kept the ${_1\over^2}\int R\CC R$ term
in the action). Note the similarities but also
the differences with the RG picture
of the critical $\phi^4$ theory below 4 dimensions
governed by a non-trivial $\CO(\epsilon\equiv 4-d)$
fixed point with a modified scaling\footnote{unlike here
where the normal scaling of the 2-point function leads
to the absence of wave-function renormalization}
and a finite number of relevant scaling fields.
We believe that although our analysis has
used crucially the simplifications inherent in the Kraichnan's
model, the above conclusions are robust enough to
explain the occurrence of multifractal exponents in
other turbulent systems, for example in the passive scalar
model with velocities correlated in time \cite{Falk3}.
Whether a similar mechanism
is responsible for the breakdown of the Kolmogorov scaling
in the full-fledged NS turbulence remains to be
seen. One of the main problems there is the lack of a small
parameter (like $\xi$ above) which will render obtaining
reliable numerical values of structure-function
exponents very difficult, if not impossible.
The situation may be different in weak turbulence
\cite{ZFL} which can provide a more gratifying test ground
for the IRG ideas.

\end{document}